\documentclass[aip,rsi,reprint]{revtex4-1}

\pdfoutput=1

\usepackage{amsmath} 
\usepackage{amssymb} 
\usepackage{textcomp} 
\usepackage{xcolor}
\usepackage[percent]{overpic} 
\usepackage{graphicx,hyperref} \hypersetup{colorlinks=true, linkcolor=blue, urlcolor=blue, citecolor=blue}
\usepackage{multirow}

\begin{document}

\newcommand{\nucl}[2]{{}^{#1}\mbox{#2}} 
\newcommand{\CNIF}{Californium Neutron Irradiation Facility (CNIF)\renewcommand{\CNIF}{CNIF}}
\newcommand{\IDL}{Instrumentation Development Laboratory (IDL)\renewcommand{\IDL}{IDL}}
\newcommand{\TTS}{Transit Time Spread (TTS)\renewcommand{\TTS}{TTS}}
\newcommand{\QE}{quantum efficiency (QE)\renewcommand{\QE}{QE}}
\newcommand{\PE}{photoelectrons (PE)\renewcommand{\PE}{PE}}
\newcommand{\IBD}{inverse beta-decay (IBD)\renewcommand{\IBD}{IBD}}
\newcommand{\NGA}{National Geospatial-Intelligence Agency (NGA)\renewcommand{\NGA}{NGA}}
\newcommand{\nuebar}{\ensuremath{\bar{\nu}_{e}}}
\newcommand{\nuebars}{\ensuremath{\bar{\nu}_{e}}'s}

\title{Invited Article: miniTimeCube}




\author{V.A.~Li}
\affiliation{Department of Physics and Astronomy, University of Hawaii at Manoa, Honolulu, HI 96822,~USA}
\author{R.~Dorrill}
\affiliation{Department of Physics and Astronomy, University of Hawaii at Manoa, Honolulu, HI 96822,~USA}
\author{M.J.~Duvall}
\affiliation{Department of Physics and Astronomy, University of Hawaii at Manoa, Honolulu, HI 96822,~USA}
\author{J.~Koblanski}
\affiliation{Department of Physics and Astronomy, University of Hawaii at Manoa, Honolulu, HI 96822,~USA}
\author{S.~Negrashov}
\affiliation{Department of Physics and Astronomy, University of Hawaii at Manoa, Honolulu, HI 96822,~USA}
\affiliation{Department of Information and Computer Sciences, University of Hawaii at Manoa, Honolulu, HI 96822,~USA}
\author{M.~Sakai}
\affiliation{Department of Physics and Astronomy, University of Hawaii at Manoa, Honolulu, HI 96822,~USA}
\author{S.A.~Wipperfurth}
\affiliation{Department of Geology, University of Maryland, College Park, MD 20742,~USA}
\author{K.~Engel}
\affiliation{Department of Geology, University of Maryland, College Park, MD 20742,~USA}
\author{G.R.~Jocher}
\email[Electronic mail:\;]{glenn.jocher@ultralytics.com}
\affiliation{Ultralytics LLC, Arlington, VA 22203,~USA}
\author{J.G.~Learned}
\affiliation{Department of Physics and Astronomy, University of Hawaii at Manoa, Honolulu, HI 96822,~USA}
\author{L.~Macchiarulo}
\affiliation{Department of Physics and Astronomy, University of Hawaii at Manoa, Honolulu, HI 96822,~USA}
\author{S.~Matsuno}
\affiliation{Department of Physics and Astronomy, University of Hawaii at Manoa, Honolulu, HI 96822,~USA}
\author{W.F.~McDonough}
\affiliation{Department of Geology, University of Maryland, College Park, MD 20742,~USA}
\author{H.P.~Mumm}
\affiliation{National Institute of Standards and Technology, 100 Bureau Dr, Gaithersburg, MD 20899,~USA}
\author{J.~Murillo}
\affiliation{Department of Physics and Astronomy, University of Hawaii at Manoa, Honolulu, HI 96822,~USA}
\author{K.~Nishimura}
\affiliation{Ultralytics LLC, Arlington, VA 22203,~USA}
\author{M.~Rosen}
\affiliation{Department of Physics and Astronomy, University of Hawaii at Manoa, Honolulu, HI 96822,~USA}
\author{S.M.~Usman}
\affiliation{Exploratory Science and Technology Branch, National Geospatial-Intelligence Agency, Springfield, VA 22150,~USA}
\affiliation{Department of Geography and Geoinformation Science, George Mason University, Fairfax, VA 22030,~USA}
\author{G.S.~Varner}
\affiliation{Department of Physics and Astronomy, University of Hawaii at Manoa, Honolulu, HI 96822,~USA}

\date{\today}

\begin{abstract}
We present the development of the miniTimeCube (mTC), a novel compact neutrino detector. The mTC is a multipurpose detector, aiming to detect not only neutrinos but also fast/thermal neutrons. Potential applications include the counterproliferation of nuclear materials and the investigation of antineutrino short-baseline effects. The mTC is a plastic 0.2\%~$\nucl{10}{B}$--doped scintillator $(13~\mathrm{cm})^3$ cube surrounded by 24 Micro-Channel Plate (MCP) photon detectors, each with an $8\times8$ anode totaling 1536 individual channels/pixels viewing the scintillator. It uses custom-made electronics modules which mount on top of the MCPs, making our detector compact and able to both distinguish different types of events and reject noise in real time. The detector is currently deployed and being tested at the National Institute of Standards and Technology (NIST) Center for Neutron Research (NCNR) nuclear reactor (20~MW$_\mathrm{th}$) in Gaithersburg~MD. A shield for further tests is being constructed, and calibration and upgrades are ongoing. The mTC’s improved spatiotemporal resolution will allow for determination of incident particle directions beyond previous capabilities.
\end{abstract}


\maketitle 

\section{Introduction: The motivation behind compact neutrino detectors and the mTC}

A number of fundamental mysteries remain in the field of neutrino physics, 
for instance the structure of the mass hierarchy of the three known neutrinos, and 
the possible existence of sterile neutrinos that interact only
through mixing.  Further, their ultimate nature as Majorana or Dirac fermions has yet to be determined.
At the same time, our understanding of neutrinos has reached
a turning point where practical applications of neutrino detection are 
becoming increasingly feasible.  This understanding, combined with recent 
developments in the areas of fast photodetectors, high-quality doped 
scintillators, electronics, and computing, have led to the possibility of a 
new generation of compact, highly instrumented neutrino detectors that were 
previously impractical and unaffordable.  These detectors will allow 
exploration of fundamental neutrino properties, as well as practical 
applications in the fields of reactor safety and nuclear security.

\begin{figure}[htbp!]
\includegraphics[width=.9\linewidth]{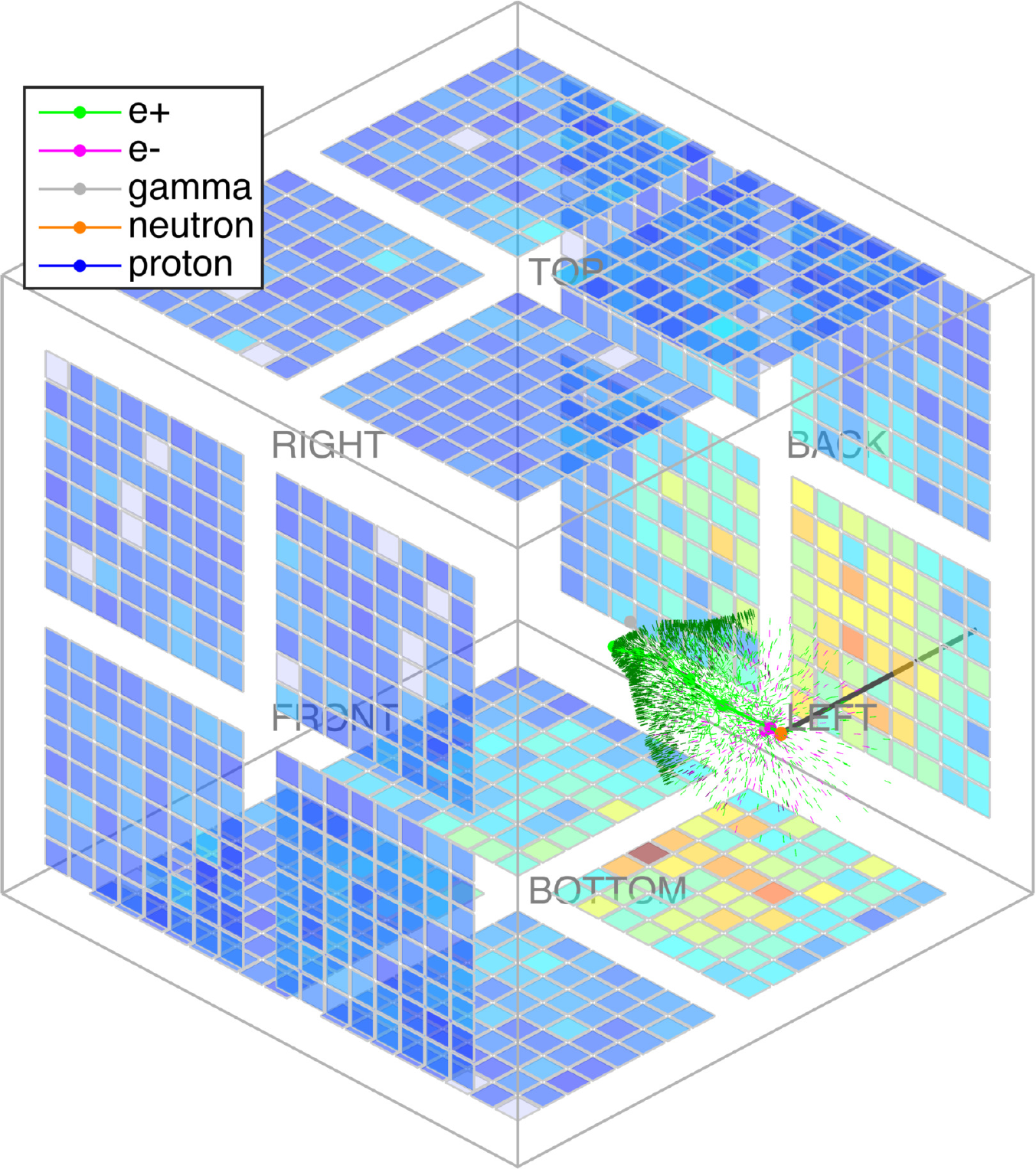}
\caption{GEANT and MATLAB simulation of a 10~MeV \nuebar~interaction in the 13~cm cubical mTC. Photons colored identically to parent particles.}
\label{fig:IBD_simulation}
\end{figure}

Specifically, measurements of reactor antineutrinos provide a number of 
scientific opportunities, such as the detailed study of neutrino
oscillations at very short baselines, and investigation of the reactor
antineutrino anomaly, which may be connected to the existence of sterile
neutrinos.\cite{Vogel:2015wua}
The miniTimeCube (mTC), shown in simulation in Fig.~\ref{fig:IBD_simulation}, 
represents a new step in this direction.
The mTC is a compact ($\sim2200~\mathrm{cm}^3$ active volume), 
densely instrumented, fast timing plastic-scintillator detector designed as a proof-of-concept for future 
reactor antineutrino detectors.  In addition to these scientific 
studies, the mTC is also designed for practical applications, such as 
directional neutrino detection and reactor monitoring for non-proliferation.

\subsection{The History and Inception of the mTC}

The motivation for a compact neutrino detector began with a study 
involving National Geospatial-Intelligence Agency (NGA), Integrity Applications Incorporated (IAI), and UH 
personnel in 2011.  It is an evolution of a CCD-based detection
concept,\cite{Learned:KamLAND_2009,Watanabe:2010} which was found to
have issues with scalability to large detection volumes.
To avoid similar problems, the mTC utilizes time as an extra 
dimension to reconstruct the event kinematics. 
In the mTC concept, a Fermat surface is defined by the first light arrival, 
leading to spatial and angular resolutions well below what one would expect 
from the scintillator decay times.\cite{Learned:2009rv}
This leads to particle location resolutions on the order of millimeters instead of 
the meter scale one would naively expect from scintillator decay time 
constants.

\subsection{Technological Context}

The mTC concept requires excellent single photon timing resolution, 
which is achieved using commercial micro-channel plate 
photomultiplier tubes with excellent intrinsic timing ($\sim 50$~ps).
Combined with readout electronics we expect single photon timing resolutions
of $100$~ps or better, corresponding to about $2$~cm spatial resolution in the 
scintillator. Further improvement is achieved by multiple pixel constraints, 
roughly scaling as $1/\sqrt{N_{\mathrm{pe}}}$.

The mTC's state-of-the-art fast-timing and pixelization allow many novel 
measurements. Although its small size may prohibit full investigation of 
some of the proposed applications, it serves as a proof-of-concept and 
model for future detectors such as NuLat.\cite{Lane:2015alq}

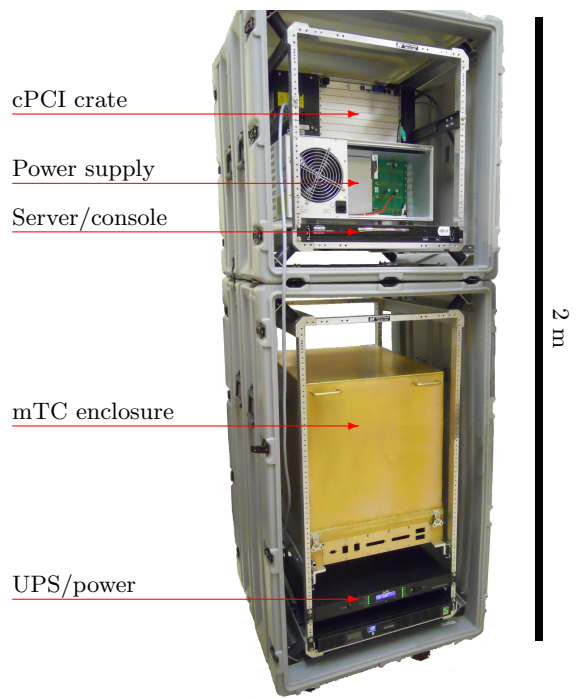
\begin{figure} 
\begin{overpic}[width=1.0\linewidth]{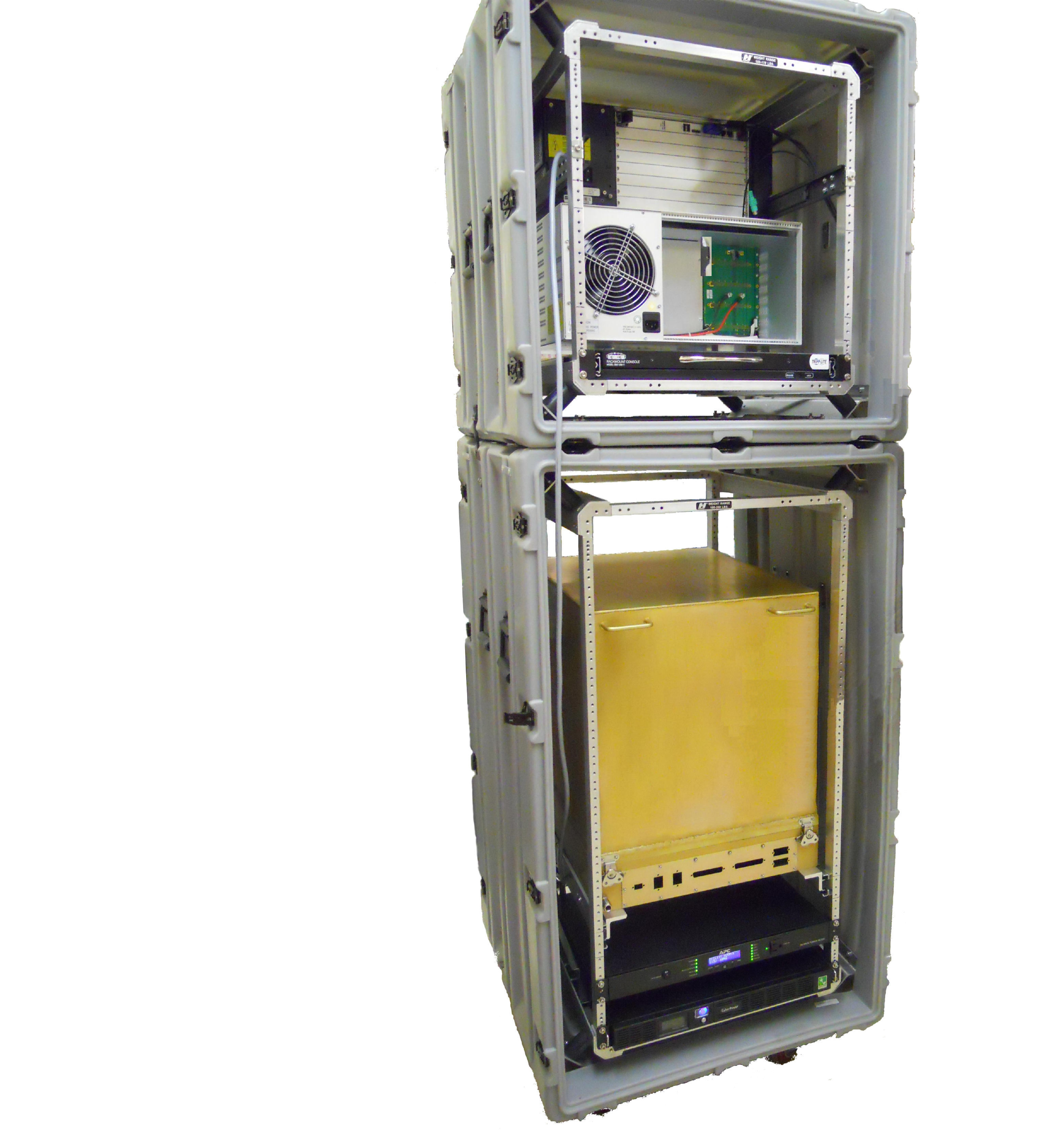}
\put(10,85){\color{red}\vector(1,0){50}}
\put(10,86){\color{black}cPCI crate}
\put(10,75){\color{red}\vector(1,0){50}}
\put(10,76){\color{black}Power supply}
\put(10,68){\color{red}\vector(1,0){50}}
\put(10,69){\color{black}Server/console}
\put(10,40){\color{red}\vector(1,0){50}}
\put(10,41){\color{black}mTC enclosure}
\put(10,15){\color{red}\vector(1,0){50}}
\put(10,16){\color{black}UPS/power}

\put(88,57){\rotatebox{-90}{2 m}}
\linethickness{3.0pt}
\put(86,9){\color{black}\line(0,1){90}}
\end{overpic} 
\caption{Photograph of mTC's mount-racks, light-tight aluminum enclosure, data acquisition system, and power supplies.}
\label{fig:mtc_racks}
\end{figure}

The preliminary design of the detector and initial performance simulations 
were conducted in 2011, with construction starting the same year. The initial
version of the detector, shown in Fig.~\ref{fig:mtc_racks}, was completed
at the end of 2013. In January 2014 we started testing and calibrating the 
detector at NIST.  A number of upgrades have been 
performed or are underway as a result of lessons learned from these 
initial studies. We expect to begin operation at the NIST reactor,
pending installation of a shielding cave to reduce neutron backgrounds,
in late 2015.

\subsection{Design of the mTC}

The core detection volume of the mTC is a 
$(13~\mathrm{cm})^3$ 
cube of plastic scintillator (Eljen Technology EJ-254), doped with 
1\% natural boron (0.2\% $^{10}\mathrm{B}$).\cite{Eljen}  The 
scintillator decay constant is $2.2~\mathrm{ns}$.

\begin{figure}[htbp!]
\includegraphics[width=1.0\linewidth]{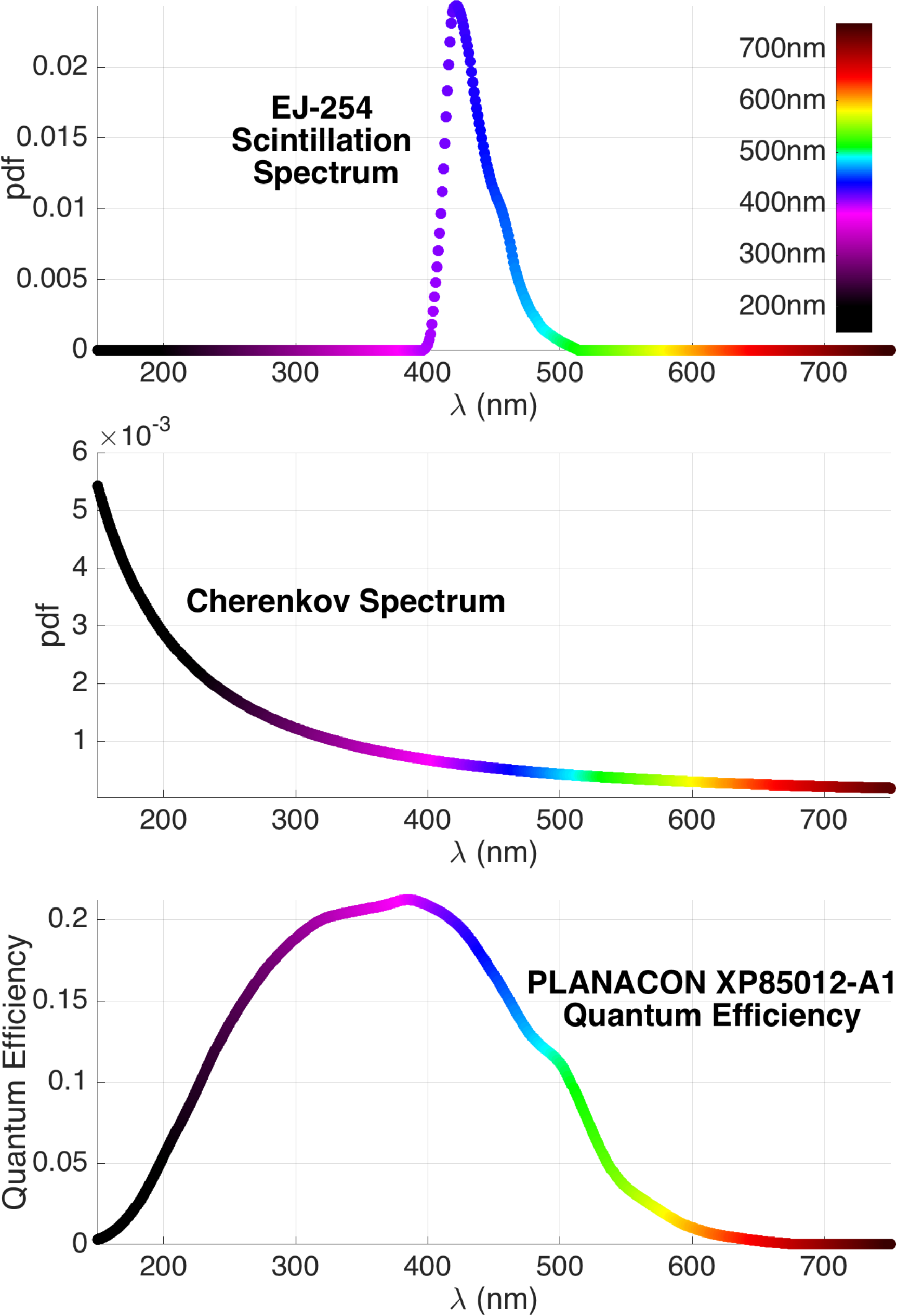}
\caption{Scintillation, Cherenkov and QE spectra for the mTC. GEANT and MATLAB MC models include all effects of chromatic dispersion. Most Cherenkov photons in the UV region attenuate very quickly and are not observed.}
\label{fig:QE_MCP}
\end{figure}

\begin{figure} [htbp!]
\begin{overpic}[width=1.0\linewidth]{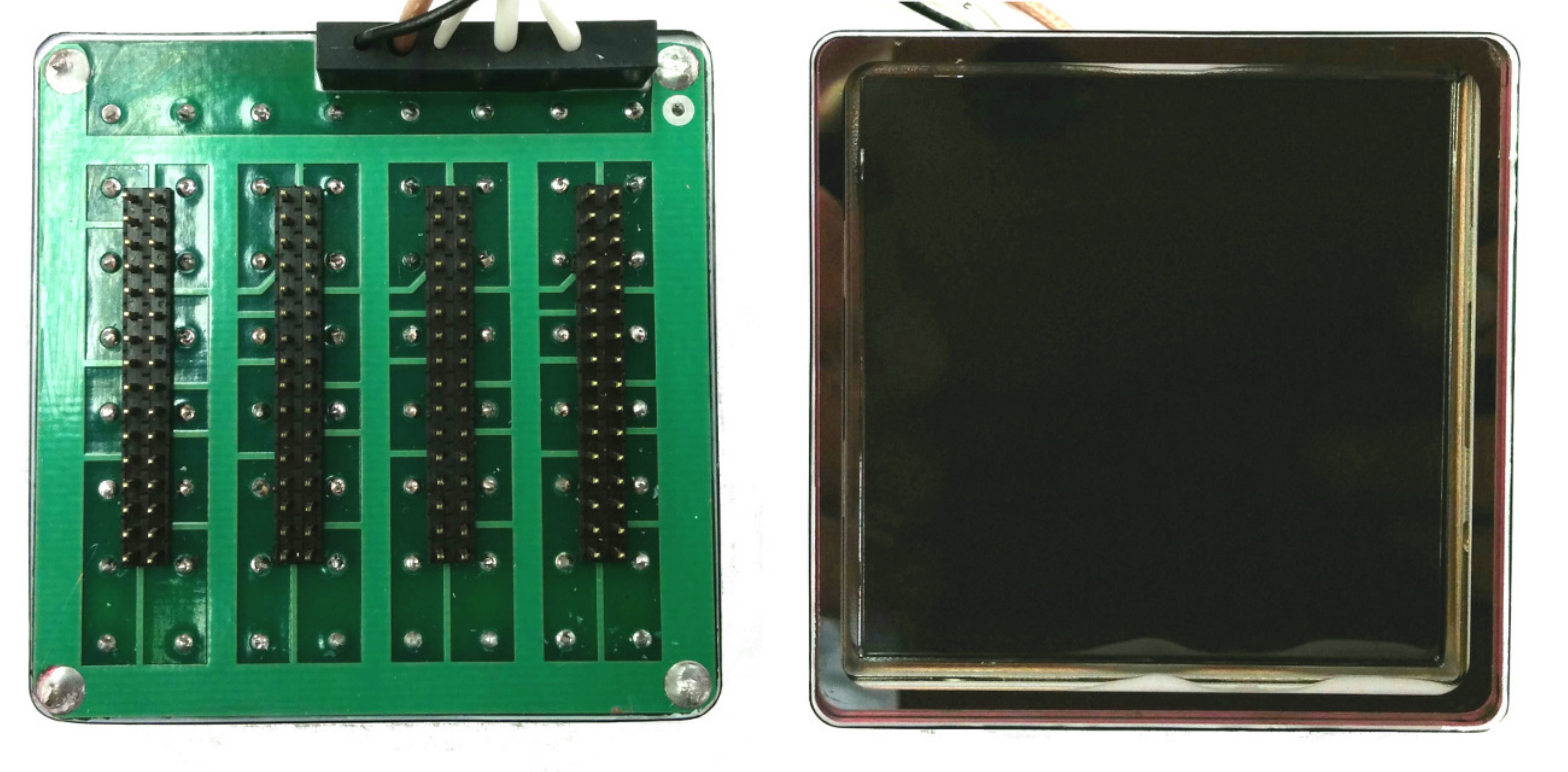}
\linethickness{3.0pt} 
\put(70,8){\color{white}59 mm}
\put(51.8,6){\color{white}\line(1,0){44.5}}
\end{overpic} 
\caption{Photograph of a PHOTONIS PLANACON MCP XP~85012, one of 24 MCPs used in the mTC.
\label{fig:MCP_Photonis_photo}}
\end{figure}

A total of 24 PLANACON MCP-PMTs (PHOTONIS XP85012), 
hereafter referred to as simply ``MCPs,'' 
shown in Fig.~\ref{fig:MCP_Photonis_photo},\cite{Photonis:2013} 
are used to detect photons from the scintillator volume.  They 
are coupled to the scintillator cube using optical grease
(Eljen Technology EJ-550), and clamped in place to the cube 
for mechanical stability. The anode plane of each MCP is 
segmented into 64 pixels, leading to a total of 1536  
readout channels.  
The scintillation and Cherenkov spectra expected for
EJ-254 are shown in Fig.~\ref{fig:QE_MCP}, along with 
the typical quantum efficiency (QE) curve of the MCP, showing
 matching of the QE to the scintillation spectrum. 
The sensitivity of the scintillator, 
including coverage factors and detection efficiency of the 
MCPs, is $\sim1000$ photoelectrons / MeV.

\begin{figure}
\begin{overpic}[width=0.8\linewidth]{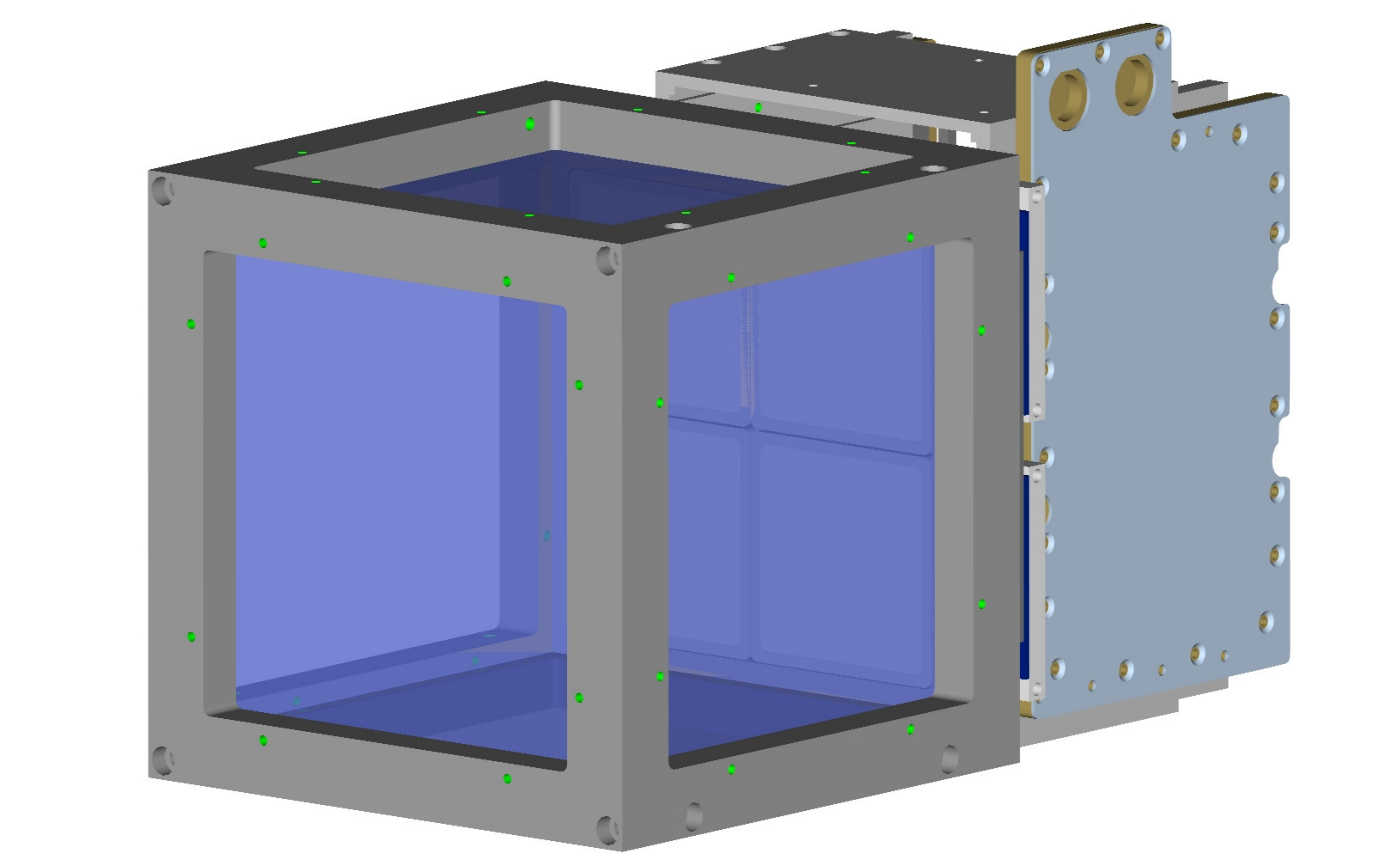}
\linethickness{3.0pt} 
\put(1,22){\rotatebox{90}{13 cm}}
\put(7,9){\color{black}\line(0,1){38}}
\end{overpic} 
\caption{
CAD of the mTC scintillator cube with one face populated with four MCPs and two electronics board stacks connected. 
\label{fig:CAD_scintillator_cube_one_face_populated}}
\end{figure}

The 1536 MCP readout channels are instrumented
with custom electronics developed at the University of Hawaii.
These electronics mount directly to the MCPs, providing 
multi-gigasample per second sampling and on-board digitization of the 
MCP signals with a timing resolution of $< 100~\mathrm{ps}$.
By preserving the excellent timing resolution of the MCPs, we retain 
the ability to study advanced reconstruction techniques (e.g., 
incorporation of the shape of the scintillator decay time distribution
and the fast timing of the Cherenkov photons).
A model of the scintillator cube with one face of 
photodetectors and corresponding readout electronics populated is shown 
in Fig.~\ref{fig:CAD_scintillator_cube_one_face_populated}.  The 
compact nature of the readout electronics keeps the 
core of the mTC compact.  The net dimensions of the cube, MCPs and 
electronics fit inside a $\sim 1/8$~m$^3$ volume. The electronics
is discussed in more detail in Section \ref{sec:Electronics}.

The main detector, ancillary electronics, and power supplies 
fit in stacked plastic cases, with a clearance footprint of 
0.75~m wide by 1.2~m deep by 2.5~m high, and requires only 115~VAC 
and a network connection for remote operation.
The assembled and integrated mTC, including associated servers for data 
acquisition, is shown in Fig.~\ref{fig:mtc_racks}.  A water-based 
chiller, with flow around 8 LPM, provides cooling needed for operation
in the shielded enclosure.  The power consumption is 
roughly 2 kW, including $\sim 1~\mathrm{kW}$ from the chiller itself.  
The size and 
power consumption make this a relatively portable detector,
capable of being operated from a truck or a ship.
 
\section{Neutrinos in mTC}

The process of identifying a neutrino interaction in the mTC is similar to the one first used in the Reines \& Cowan experiment~\cite{Cowan:1992xc,Reines:1956rs} in 1956, and used by many experiments since.

An electron antineutrino emitted from a nuclear reactor interacts with a proton inside the plastic scintillator medium, producing a positron and a neutron via \IBD:
\begin{equation} \label{eq:IBD}
\bar{\nu}_e + p \to e^+ + n
\end{equation}

This reaction has a cross-section of $\sigma_{tot} \cong 5\times10^{-43}\ \mathrm{cm}^2$ at a neutrino energy $E_\nu = 2\ \mathrm{MeV}$ and an energy threshold of $E_\nu = 1.806$ MeV (in the lab frame, where the proton is at rest). The characteristic time-scale between {\em prompt} (positron annihilation) and {\em delayed} (neutron capture) signals is used as the primary signature for identifying neutrino events. As outlined below, the positrons scatter nearly isotropically after the neutrino interaction, with the positron taking most of the kinetic energy and the neutron taking most of the momentum. If one records the direction and energy of the positron and the first scatter of the neutron, one can back-reconstruct the incident direction of neutrino. Additionally, further scatters of the neutron can also be used to improve the reconstruction.

\subsection{Prompt Signal}

\begin{figure}[htbp!]
\includegraphics[width=1.0\linewidth]{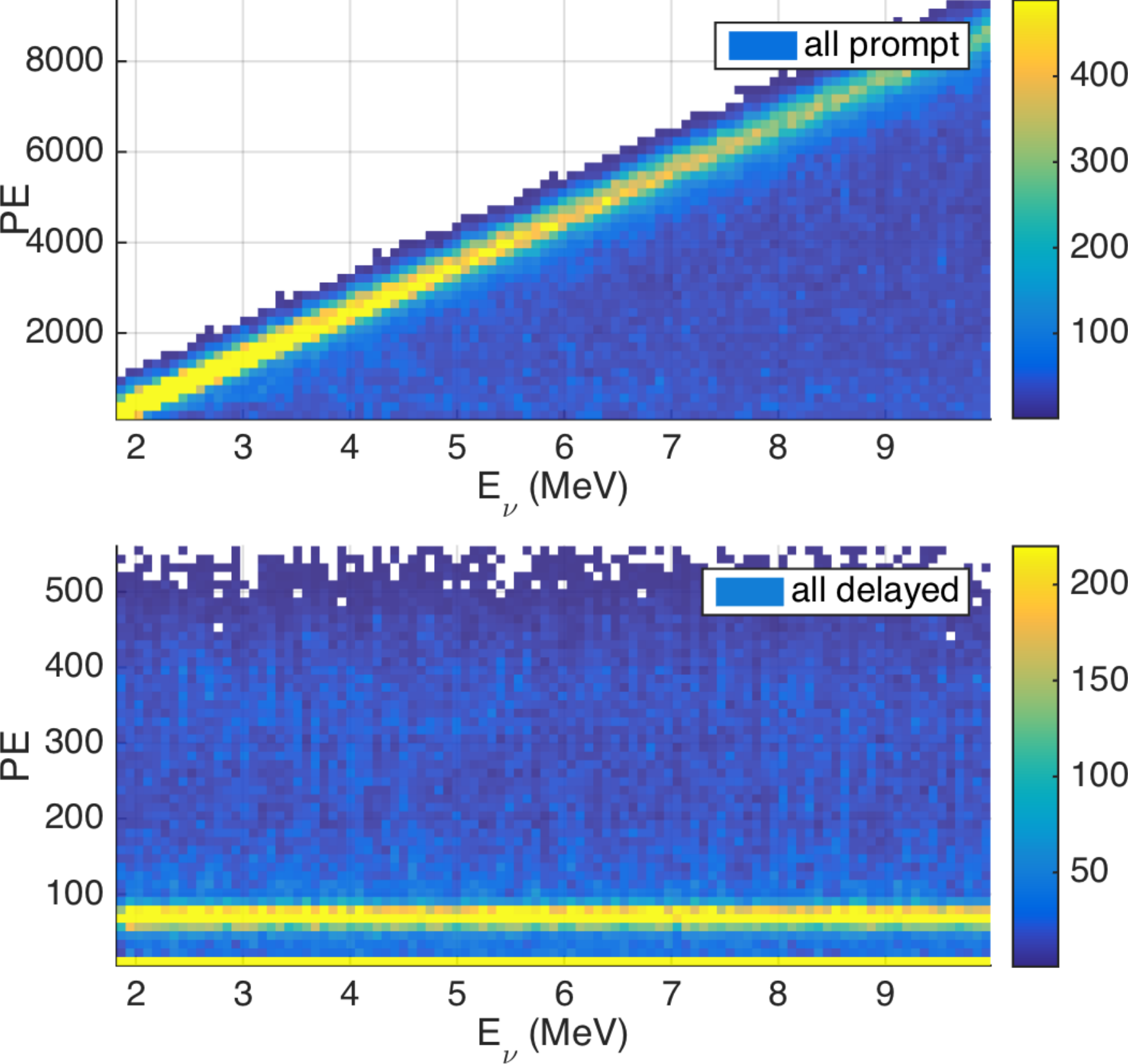}
\caption{Simulated distributions of the number of PE produced as a function of \nuebar~energy for the prompt signal (top) and delayed signal (bottom). In top figure, a long tail of under-estimated energies is produced by longer positrons leaving the detector. In the bottom figure, the long tail of higher energy delayed events is due to the 478 keV gamma produced on neutron capture depositing part of it’s energy randomly inside the detector.}
\label{fig:PE_dist}
\end{figure}

\begin{figure}[htbp!]
\includegraphics[width=1.0\linewidth]{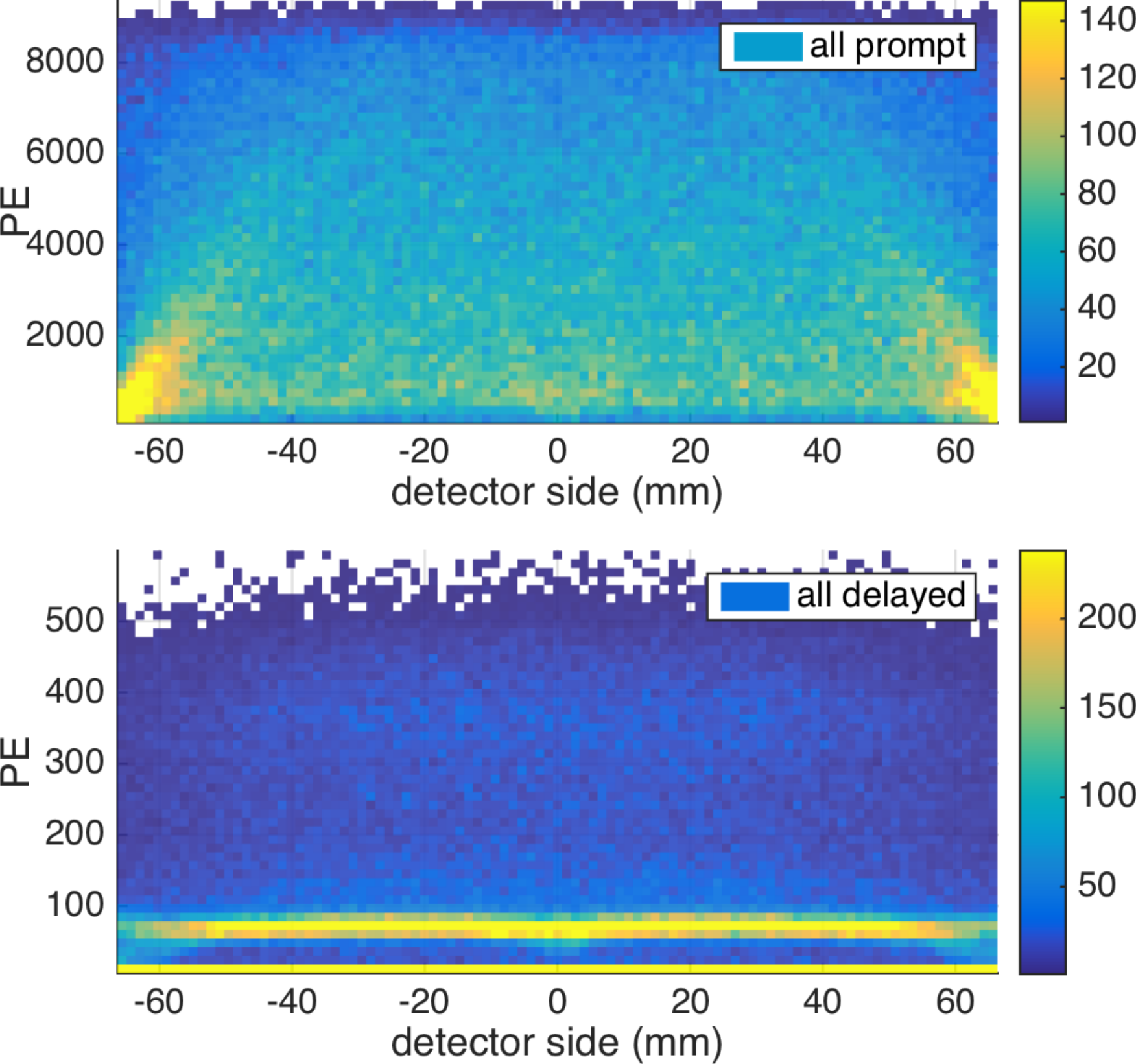}
\caption{Simulated distributions of the number of PE produced for the prompt signal (top) and delayed signal (bottom) as a function of vertex location.   0~mm is the center of the detector, and 67~mm is at the edge.}
\label{fig:glennf11v}
\end{figure}

The IBD prompt signal generates anywhere from several hundred to several thousand Photo-Electrons (PEs) in the mTC, as shown in Figs.~\ref{fig:PE_dist} and \ref{fig:glennf11v}. The energy of this signal is used to reconstruct the incoming \nuebar ~energy, and the location of the signal may be used for directional determination of the \nuebar ~angle, however weakly, by pairing it with the delayed signal location. The prompt signal is composed of a short positron track ($\sim 1~\mathrm{cm}$), any electrons it may interact with, and two equal and opposite 511~keV gammas produced upon positron annihilation.

Simulation projects that the mTC reach about 10-15\% \nuebar~energy resolution, ultimately limited by 
its small size. The positron track at the lower range of the \nuebar~energy spectrum is on the order of a cm, though higher energy \nuebars~will produce longer tracks proportional to their energy above the $E_{\bar{\nu}_e} > 1.8~\mathrm{MeV}$ threshold. As a result, higher energy \nuebars~tend to produce positrons which leave the detector with ever greater likelihood, causing a certain amount of energy under-estimation at higher \nuebar~energies.

In addition to severed positron tracks, a second problem arises at all \nuebar~energies: uncertainty in prompt signal is introduced via the two 511~keV gammas. In a larger detector such as KamLAND, the annihilation gammas typically deposit their full energy within the scintillation volume. In a very small detector like mTC, these gammas deposit varying amounts of energy from event to event, smearing the prompt energy resolution. On average the annihilation gammas deposit about 1/3 of their energy in the mTC, but the proportion varies event to event, and is impossible to predict {\it a priori} for a specific IBD event.

More information on expected energy resolutions can be found in Section \ref{sec:Reconstruction}.

\subsection{Delayed Signal}

\begin{figure}[htbp!]
\includegraphics[width=1.0\linewidth]{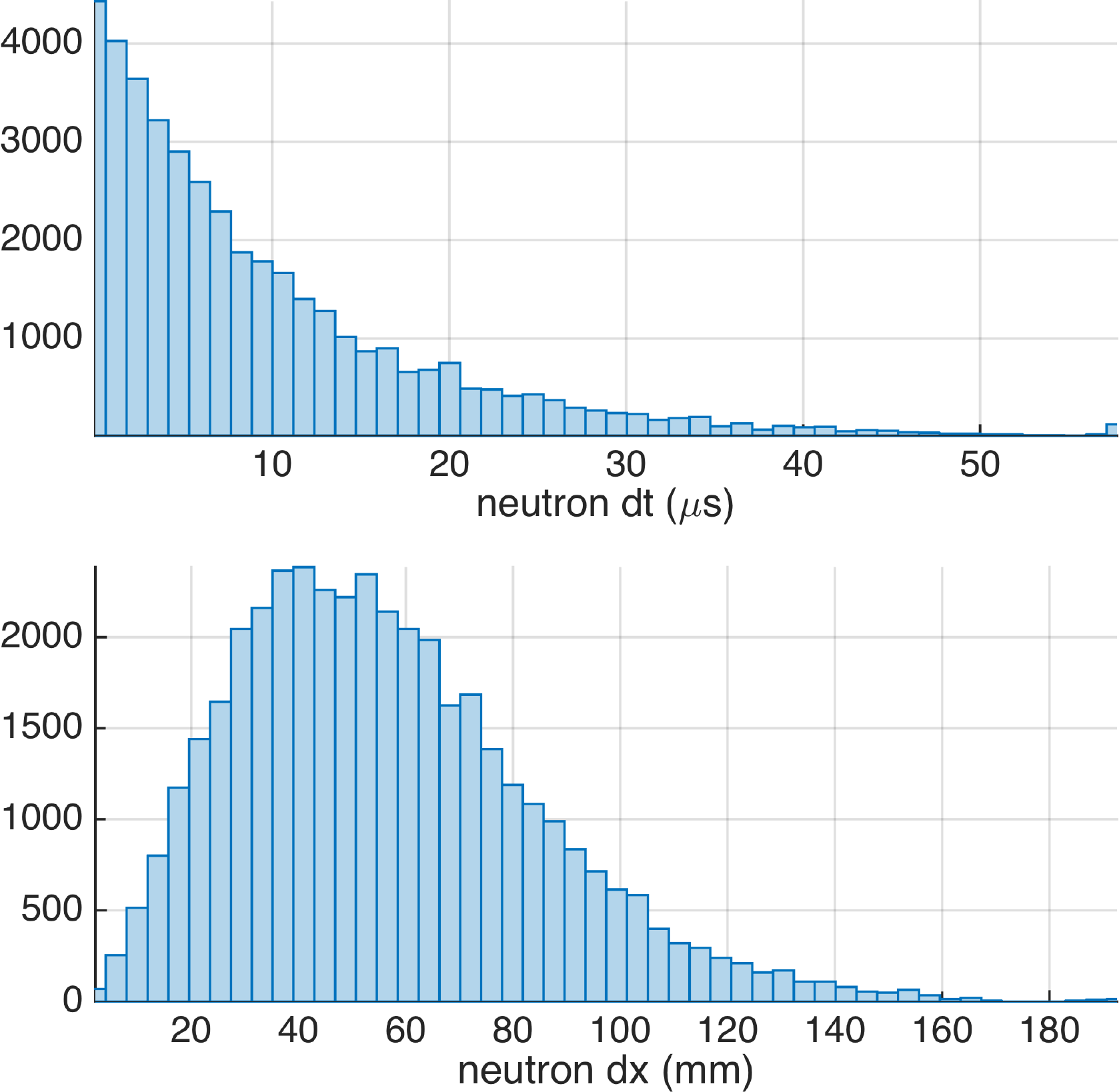}
\caption{Simulated frequencies, in arbitrary units, of Monte Carlo generated IBD events as a function of time and distance of neutron production to neutron capture in the mTC scintillator.}
\label{fig:glennf1v}
\end{figure}

The neutron from the neutrino interaction scatters elastically on the scintillator medium and, after thermalizing, captures on the $\nucl{10}{B}$ embedded in the scintillator. On average, the neutron travels for a few centimeters before being captured, as shown in Fig.~\ref{fig:glennf1v}.

The primary reaction for neutron capture is:\cite{Knoll:2000fj,Pawelczak:2014,Wright:2010bu}
\begin{equation} \label{eq:capture_B10}
\begin{split}
 n + \nucl{10}{B} \to & \nucl{7}{Li} (1015\; \mathrm{keV}) + \nucl{4}{He} (1775\; \mathrm{keV}), \quad \sim 6 \% \\
		 \to & \nucl{7}{Li}^* + \nucl{4}{He} (1471\; \mathrm{keV}), \qquad \qquad \quad \sim 94 \% \\
		  & \hookrightarrow \nucl{7}{Li}^* \to \nucl{7}{Li} (839\; \mathrm{keV}) + \gamma (478\; \mathrm{keV})
\end{split}
\end{equation}

The cross-section  for neutron capture on $\nucl{10}{B}$ as well as the linear attenuation coefficient as a function of neutron energy have been studied.\cite{Eljen,Wright:2010bu,Crane:1991,ENDF_web}  For a completely thermal neutron ($E_n = 0.025$ eV), the total cross-section is equal to 3836~barns. 
The neutron from an IBD reaction can also be captured on a proton in $\sim$\;180~$\mu$s, resulting in 2.2~MeV $\gamma$'s depositing energy via Compton scattering.\cite{Agashe:2014kda}  The fraction of thermal neutrons captured on $\nucl{10}{B}$ is 25.6 times larger than on $\nucl{1}{H}$.\cite{Eljen}

To confirm our understanding of the Monte Carlo results shown in Fig. \ref{fig:glennf1v}, we can estimate the detector's neutron-capture efficiency analytically.  Because the majority of neutrons capturing on $\nucl{10}{B}$ is so large, neutron thermalization and capture on $\nucl{10}{B}$ is the dominating process in determining the detector's neutron-capture efficiency. We can therefore get a rough estimate of this number by considering a typical neutron undergoing this process. For this calculation, we treat the neutron's path as a random-walk series of elastic scatters on $\nucl{1}{H}$ in two parts: 1) production to thermalization, and 2) thermalization to capture. Combining these two results will give us a general idea of where neutrons are likely to be lost and therefore what fraction of neutrons should capture without escaping the cube.

We will take a typical IBD neutron to have $K_{init}~\sim~4~\text{keV}$. At these neutron energies, we can safely use nonrelativistic kinematics: $K \sim \mathcal{O} (1~\text{keV}) \ll m_0$.

1) Production to thermalization, $d_{therm}$:\\
\begin{equation}
d_{therm} \sim \lambda_{es} \sqrt{N_{therm}},
\end{equation}
where $\lambda_{es}$ is the mean free path (MFP) for elastic scattering and $N_{therm}$ is the number of steps to thermalization.
We can get $\lambda_{es}$ from the cross-section $\sigma_{es}$ (20 barns for hydrogen) and the volume density of targets $n_H$:
\begin{equation}
\lambda_{es}\ \sim\ \frac{1}{n_H\ \sigma_{es}}
\end{equation}
To find $N_{therm}$, we assume that on average the neutron loses half its excess KE on each collision with a proton. Then $N_{therm}$ is simply: 
\begin{equation}
N_{therm} \sim \log_{2} \bigl( \frac{K_{init}}{K_{therm}} \bigr)
\end{equation}

Combining the above equations and data from the scintillator manufacturer,\cite{Eljen} we get $\sim17$ steps at $\sim1~\text{cm}$ each for a distance of:
\begin{equation}
 d_{therm}\ \sim\ 4~\text{cm}
\end{equation}

2) Thermalization to capture, $d_{cap}$:\\
Once thermalized, the neutron will typically capture after traveling the corresponding MFP, $\lambda_{cap}$. Because this distance is longer than $\lambda_{es}$ above, the neutron will continue its elastic scattering on H during this time, with a number of steps equal to $\lambda_{cap}/\lambda_{es}$. The displacement for this part of the process is therefore:
\begin{equation}
   d_{cap}\ \sim\ \lambda_{es} \sqrt{\frac{\lambda_{cap}}{\lambda_{es}}}\ \sim\ \lambda_{es} \sqrt{ \frac{1}{\lambda_{es}} \cdot \frac{1}{n_{B10}\ \sigma_{cap}} }\ \sim\ 1.5~\text{cm}
\end{equation}

3) Total (production to capture), $d_{tot}$:\\
Finally, taking $d_{therm}$ and $d_{cap}$ to be two steps of a random walk, we have:
\begin{equation}
d_{tot} \ \sim \ \langle d \rangle \sqrt{N} \ = \ \frac{d_{therm} + d_{cap}}{2} \sqrt{2}\ \sim 4~\text{cm},
\end{equation}
as the typical total distance between production and capture (again, ignoring corrections for effects like capture before thermalization, capture on hydrogen, etc.).

4) Neutron-capture Efficiency, $n_{captured}/n_{total}$:\\
We can use an imaginary sphere of radius $d_{tot}\sim4~\text{cm}$ to roughly estimate the capture rates in various regions of the cube. For example, a neutron produced at the surface of the cube but near the center of a face will generally have $\sim1/2$ probability to move inward and capture or to move outward and escape; however, a neutron produced deeper than $d_{tot}\sim4~\text{cm}$ into the face will most likely capture inside the cube. Averaging over the depth indicates that $\sim3/4$ of the neutrons produced in this region should capture inside the cube. We can make similar estimates for the rates in the other regions of the cube (i.e., edges, corners, and interior) and combine these estimates to get our final result:
\begin{equation}
\frac{n_{captured}}{n_{total}}\ \sim\ \frac{1}{2}
\end{equation}

This is in general agreement with simulations indicating that $\sim55\%$ of the neutrons produced inside the cube capture without escaping.\\

Using this same approach, we can get a rough estimate for how long this process might take: $t_{tot} = t_{therm} + t_{cap}$. It is relevant to keep in mind that $N_{therm}$ depends on $K_{init}$ as discussed above.

1) Production to thermalization, $t_{therm}$:\\
As in the distance calculation, we will assume that on average, the neutron loses half of its kinetic energy on each collision. (For this calculation, we approximate $K_{therm}~\sim~0$). After $n$ collisions, this becomes $K_n~=~2^{-n}~K_{init}$. We then immediately have:
\begin{equation}
 v_n\ =\ \sqrt{\frac{2K_n}{m}}\ =\ 2^{-n/2}\sqrt{\frac{2K_{init}}{m}}\ =\  2^{-n/2}\ v_{init}
\end{equation}
Since the distance travelled in each step is $\lambda_{es}$ from above, the time for each step is:
\begin{equation}
 t_n\ =\ \frac{\lambda_{es}}{v_n}\ =\ 2^{n/2} \frac{\lambda_{es}}{v_{init}}
\end{equation}
The total time to thermalization is then the sum of these steps over $N_{therm}$ terms:
\begin{equation}
 t_{therm}\ \sim\ \sum t_n\ =\ \sum_{n=0}^{N_{therm}} 2^{n/2}\ \frac{\lambda_{es}}{v_{init}}\ \sim\ 10~\mu\text{s}
\end{equation}

2) Thermalization to capture, $t_{cap}$:\\
The average speed after thermalization is constant by definition, so the time to capture will be $\lambda_{cap}$ from above divided by this speed:
\begin{equation}
 t_{cap}\ \sim\ \frac{\lambda_{cap}}{v_{N_{therm}}}\ =\ 2^{N_{therm}/2} \frac{\lambda_{cap}}{v_{init}}\ \sim\ 10~\mu\text{s}
\end{equation}

3) Total (production to capture), $t_{tot}$:\\
Combining these results gives us:
\begin{equation}
 t_{tot}\ =\ t_{therm} + t_{cap}\ \sim\ 20~\mu\text{s},
\end{equation}
which is also in general agreement with the Monte Carlo.

Reconstruction of the neutrino's direction largely depends on the neutron direction reconstruction, improved by the positron direction and energy.  Having one or more neutron scatters improves the resolution, but even the neutron capture location after many scatters retains information on the initial neutron direction, as was demonstrated in the CHOOZ experiment.\cite{Abe:2014bwa}  Full reconstruction algorithms, currently under development, will take all the information into account in solving for incoming neutrino direction.

The light yield of these neutron scatters can present some difficulty when detecting and reconstructing events. Ionization density quenching on two charged particles ($\nucl{4}{He}$ and $\nucl{7}{Li}$) with $\gtrsim 2.3$~MeV kinetic energy in the reaction, Eq.~\eqref{eq:capture_B10}, results in a small total light output, about 60~keV electron-equivalent energy deposition.\cite{deMeijer:2005zz} However, due to the small size  of the detector and high MCP surface coverage, mTC has the high light collection efficiency crucial to detect the weak light from these delayed signals. As a result, only in relatively small-volume ($\sim$Liter sized) $\nucl{10}{B}$-doped scintillator detectors can incident antineutrino direction currently be reconstructed based on neutron directionality.\cite{deMeijer:2005zz}  

\subsection{mTC at NIST Reactor}

\begin{figure}[htbp!]
\begin{overpic}[width=1\linewidth]{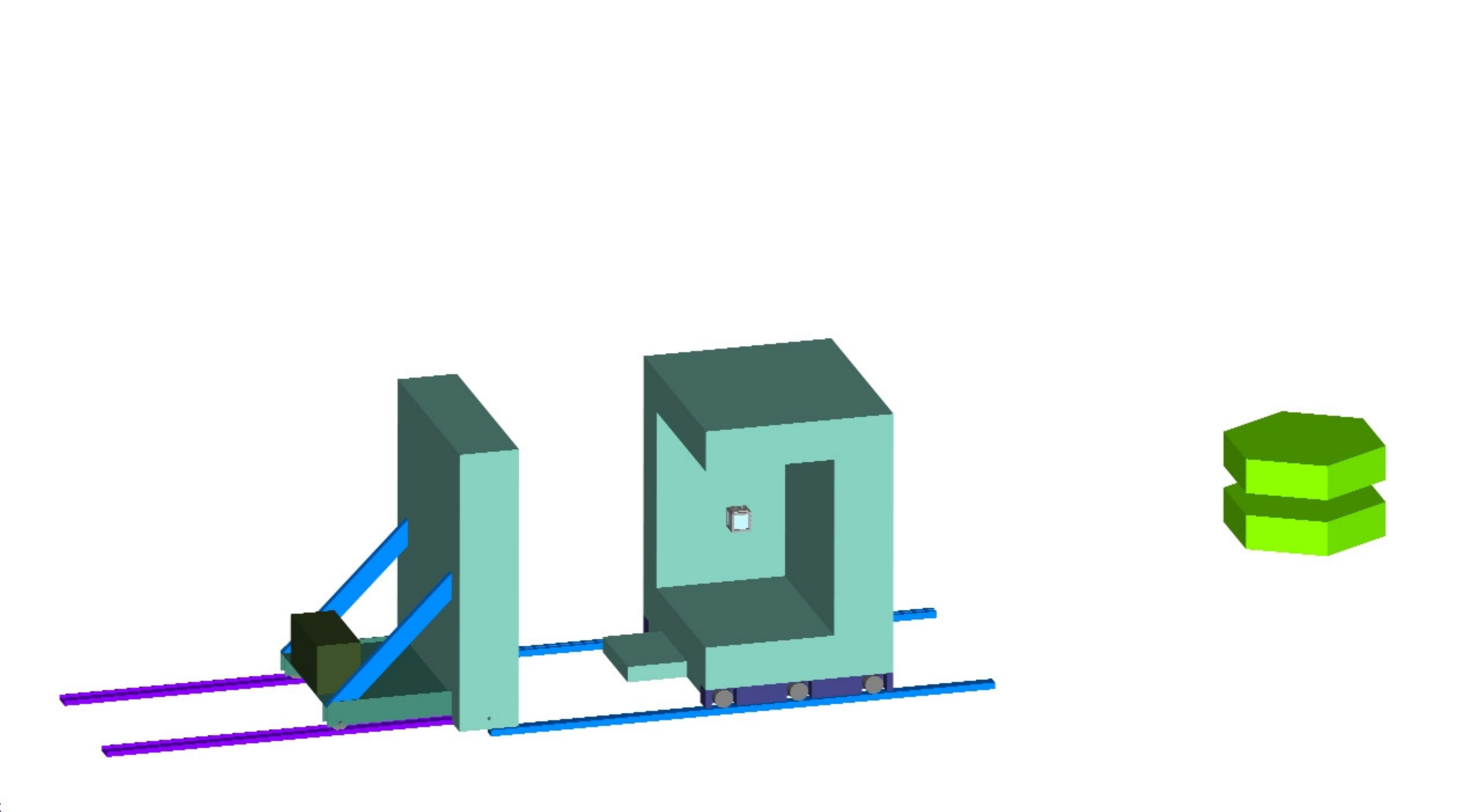}

\put(30,45){\color{red}\vector(0,-1){15}}

\put(50.5,45){\color{red}\vector(0,-1){25}}
\put(44,46){\color{black}mTC scintillator}

\put(30,45){\color{red}\vector(1,-1){15}}
\put(19,46){\color{black}shielding}

\put(89,45){\color{red}\vector(0,-1){20}}
\put(78,46){\color{black}nuclear fuel}

\linethickness{3.0pt} 
\put(70,3.5){\color{black}$\sim$\;5 m}
\put(51,2){\color{black}\line(1,0){38}}
\end{overpic}
\caption{Relative location of the scintillator cube inside the movable cave (one face made transparent) with respect to the reactor core (upper and lower fuel segments are approximated by two hexagonal prisms).}
\label{fig:mTC_NIST_reactor_CAD}
\end{figure}

\begin{figure}[htbp!]
\includegraphics[width=1.0\linewidth]{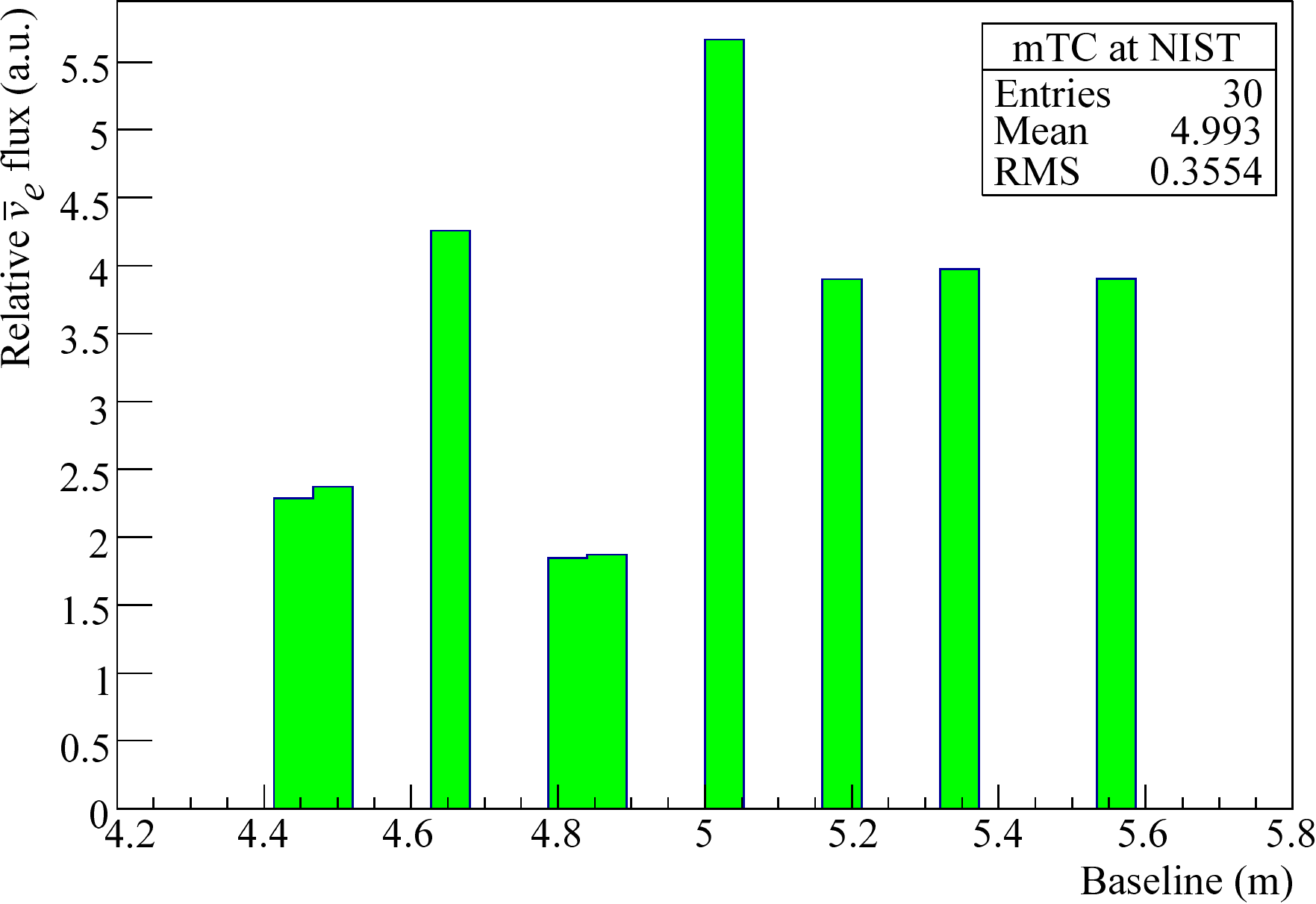}
\caption{Relative distribution of $\bar\nu_e$ flux as a function of baseline from a nominal mTC position to each fuel element in the core.   %
The mean source location of flux is at $\sim 5$~m and the effective spread is 0.36~m, or an inherent smearing of about 7~\% on the baseline.  Specifics of this distribution will vary by fuel loading conditions.
\label{fig:NIST_baseline}}
\end{figure}

The mTC currently sits on-site at the NIST Center for Neutron Research (NCNR), which houses the NIST 20 MW$_\mathrm{th}$ split-core research reactor. This reactor has a compact core (Figs.~\ref{fig:mTC_NIST_reactor_CAD}--\ref{fig:NIST_baseline}) with 30 fuel elements, each containing 2 segments of highly-enriched uranium fuel $\mathrm{U}_3\mathrm{O}_8$/Al ($\nucl{235}{U}$, 93\% enrichment). Fuel elements are submerged in heavy water which serves as a moderator and coolant. The upper and lower fuel segments, each 27.9 cm high, are separated by a 17.8 cm unfueled gap which serves as a ``flux trap'' to minimize the fast-neutron and gamma backgrounds in the neutron beam lines.  The overall dimensions of the core are 1.12 m in diameter by 0.74 m in height. The NIST reactor cycle is 38 days on followed by 10 days off for refueling.  

Full Monte Carlo N-particle (MCNP) simulations of the core are available to onsite collaborations.\cite{Hanson:2004,Hanson:2011}

\begin{figure}[htbp!]
\includegraphics[width=1.0\linewidth]{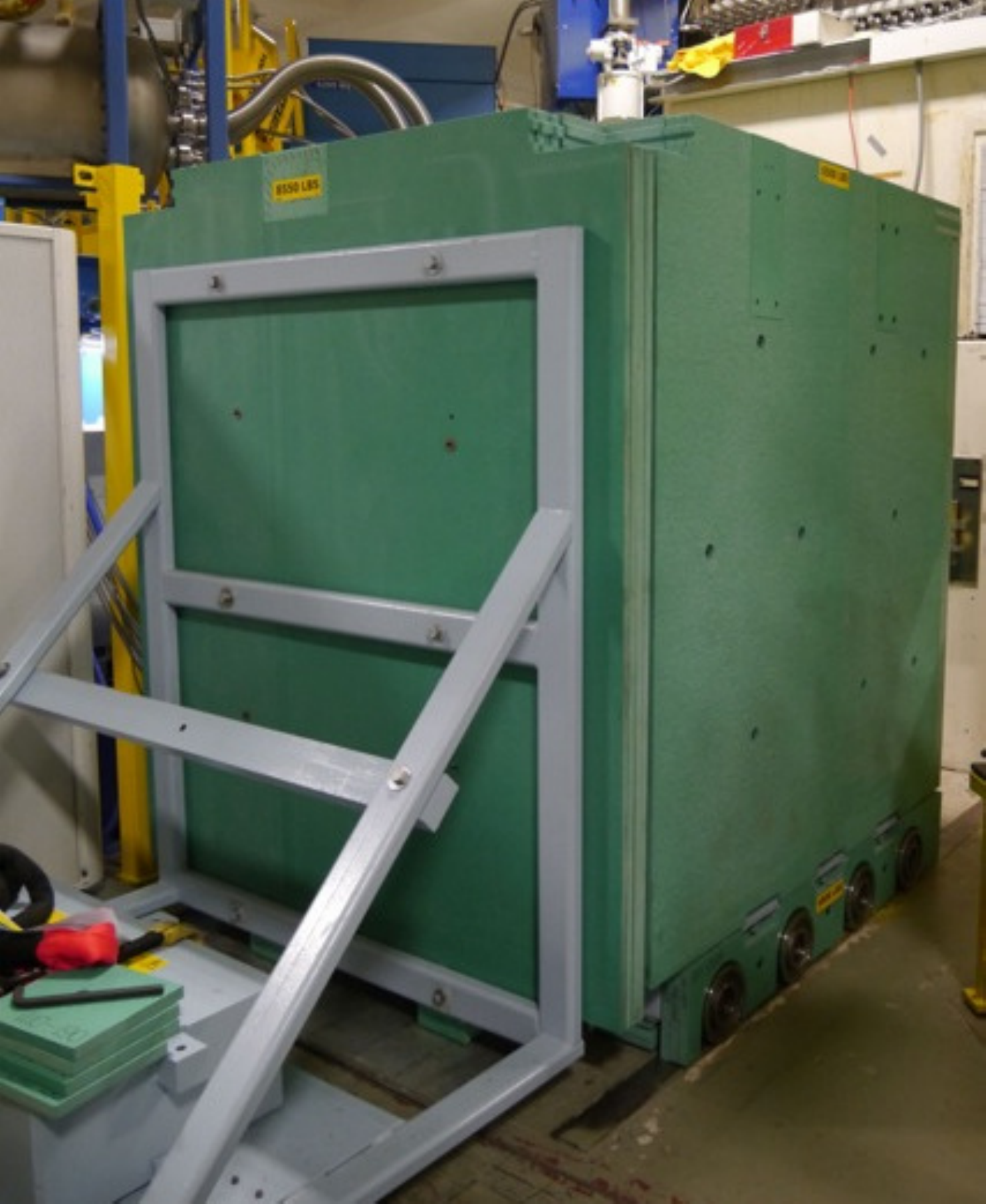}
\caption{Photograph of the mTC shielding next to the reactor.}
\label{fig:Photo_NIST_reactor_and_CAD_mTC_cave_3}
\end{figure}

Using a total thermal power of 20~MW$_\mathrm{th}$, an average number of $6\ \bar\nu_e$ produced per fission (with $\sim$~1.5~$\bar\nu_e$ above IBD threshold),  and thermal energies released per fission of $\nucl{235}{U}$ and $\nucl{238}{U}$, one can roughly estimate the total number of neutrinos produced at the reactor core to be $\sim 4 \times 10^{18}\ \mathrm{s}^{-1}\ \bar\nu_e$.
This corresponds to a flux of $\sim 1.1 \times 10^{12}\ \mathrm{cm}^{-2}\; \mathrm{s}^{-1}\ \bar\nu_e$ at the miniTimeCube location $\sim$ 5 m away from the center of the reactor core, Fig.~\ref{fig:Photo_NIST_reactor_and_CAD_mTC_cave_3}. It further corresponds to a number of antineutrino interactions with $\nucl{1}{H}$ via IBD reaction in the plastic scintillator on the order of a few events per day. 

More precisely, the total number of expected antineutrinos from the reactor observed in the detector is given by
\begin{equation} 
\begin{split}\label{eq:rate}
 N_{\bar\nu_e}^{\mathrm{obs}} = 
 \frac{N_p}{4 \pi L^2} \int \epsilon_{det} P(\bar\nu_e \to \bar\nu_e) 
 \frac{ d \sigma (E_{e^+}, E_{\bar\nu_e}) }{ d E_{e^+} } & \times \\
 \frac{ d ^2 N_{\bar\nu_e} (E_{\bar\nu_e}, t) } { d E_{\bar\nu_e} d t} d E_{e^+} d E_{\bar\nu_e} d t , &
\end{split}
\end{equation}

\noindent where $N_p$ --- number of hydrogen atoms in the scintillator (``free protons''); $L$ --- distance between production and detection points of the antineutrino,  $\epsilon_{det}$ --- detector efficiency $\sim$~30~\% based on MC simulations for IBD detection in the mTC; $P(\bar\nu_e \to \bar\nu_e)$ --- survival probability of electron antineutrino;\cite{Agashe:2014kda,Lane:2015alq} and $ \frac{ d \sigma (E_{e^+}, E_{\bar\nu_e}) }{ d E_{e^+} }$ --- differential cross-section of the IBD process as a function of positron energy $E_{e^+}$ and antineutrino energy $E_{\bar\nu_e}$.\cite{Vogel:1999zy}

Further details on antineutrino production at reactor facilities can be found in the literature, including: fuel time-dependence for the NIST nuclear reactor,\cite{Hanson:2004,Hanson:2011} evaluation of thermal energies released per fission of the four main isotopes,\cite{Kopeikin:2004cn} and spectrum of antineutrinos produced from the four main isotopes.\cite{Dwyer:2014eka,Mueller:2011nm}

In addition to NIST, we have actively considered two other deployment sites: Typical Power Reactors (TPR) and nuclear-powered ships.
Their parameters are listed in Table~\ref{table:mTC_sites}.

\begin{table}[h]
 \caption{Approximate parameters at potential mTC deployment sites, including NIST, a typical power reactor, and a nuclear-powered ship. ``Compact core'' indicates a core where all fuel elements are contained within a few meter radius. \label{table:mTC_sites}}
 \begin{tabular}{| c || c | c |}
 \hline
     Parameter & NIST & TPR \\\hline \hline
     Power, GW$_{\mathrm{th}}$ & 0.02 & 3\\
     $\langle$Baseline$\rangle$, m & 5 & 25\\
     Fuel & HEU & mixed\\
     Fuel cycle, on/off days & 38/10 & 400/10 \\
     Compact core & $\Box$\!\!\!\!\checkmark & $\Box$ \\
     $\langle$Event rate$\rangle$, $\bar\nu_e /$ day & $\sim 1$ & $\sim 10$ \\
\hline
\end{tabular}
\end{table}

\section{Backgrounds}

Backgrounds in the mTC come from several sources.  First there are the ``natural sources'', most prominently cosmic radiation.  Of those, which consist of high energy neutrons, gammas, and muons, along with their collisional products, potentially the most serious for IBD detectors is the nearly irreducible background of some long lived muon-produced isotopes as we discuss below in Section A. The local environmental backgrounds, such as radioactivity, are not as much of a problem as the cosmic ray associated backgrounds. In Section B we discuss backgrounds relevant to a reactor and specifically the NIST reactor location.

\subsection{Cosmic Ray Backgrounds}

\begin{figure}
\includegraphics[width=1.0\linewidth]{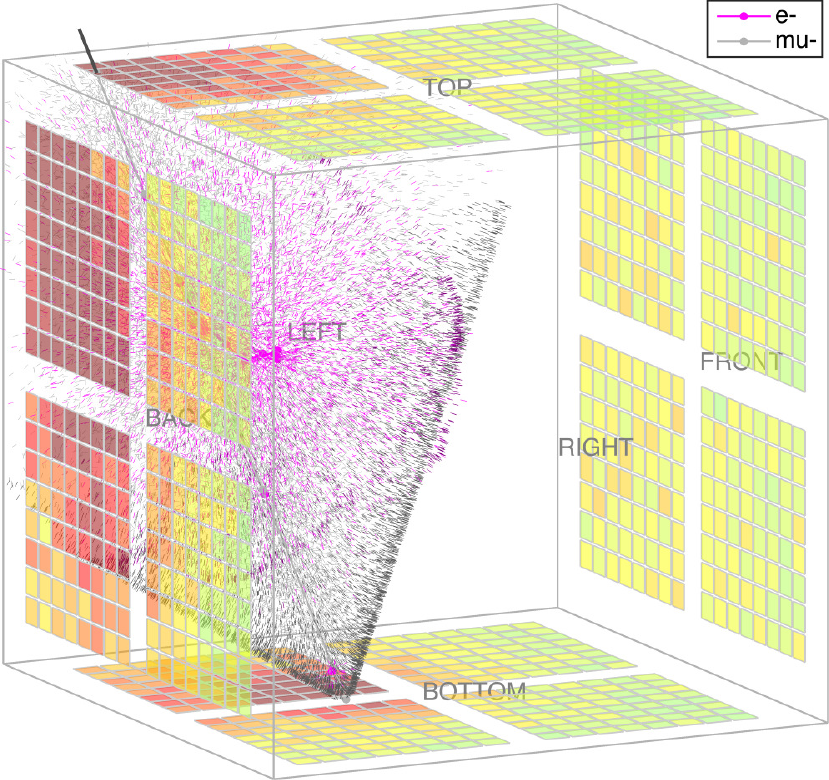}
\caption{A simulated muon traversing the mTC, with scintillation photons and Cherenkov cone visible.
\label{fig:muon_simulation}}
\end{figure}

Cosmic rays produce an inescapable background for IBD detectors.  Unfortunately, all the reactors to which we may have access are at best a few meters water equivalent (mwe) under the surface.  About 2~mwe is enough to shield from extensive air showers, clearing the remnant hadrons and most electromagnetic components.  Muons, however, penetrate to the greatest depths, in ever decreasing numbers but increasing mean energies.  These muons may generate local particles, and so shielding is somewhat of a double edged sword.  Sea-level muons make neutrons and other hadrons in nuclear interactions, though with something on the order of a 2~km radiation length.  The mean muon energy at the Earth's surface is about 2 GeV with a penetrating power of about 10 mwe.  Muons coming through the mTC (at about 1/s) often ($\sim$10\%) come with knock-on electrons (Fig.~\ref{fig:muon_simulation}).  More dangerous are  gammas and neutrons, which can fake the prompt signature of a neutrino. 

Precise calculations of these rates are difficult because they depend upon details of the overburden, the local geometry, and shielding in particular. Isotopes and various spallation products of cosmic-ray muons can be a serious background for neutrino signals. Although many of the isotopes can be filtered from analysis using various cuts, long-lived isotopes such as $\nucl{8}{He}$ and $\nucl{9}{Li}$ may have lifetimes on the order of a second and decay by beta emission into neutron-unstable daughters. These are two backgrounds that can mimic IBD events in the mTC, but are in fact negligible, as we show below.

\begin{figure}
        \begin{center}
                \includegraphics[width=1.0\linewidth]{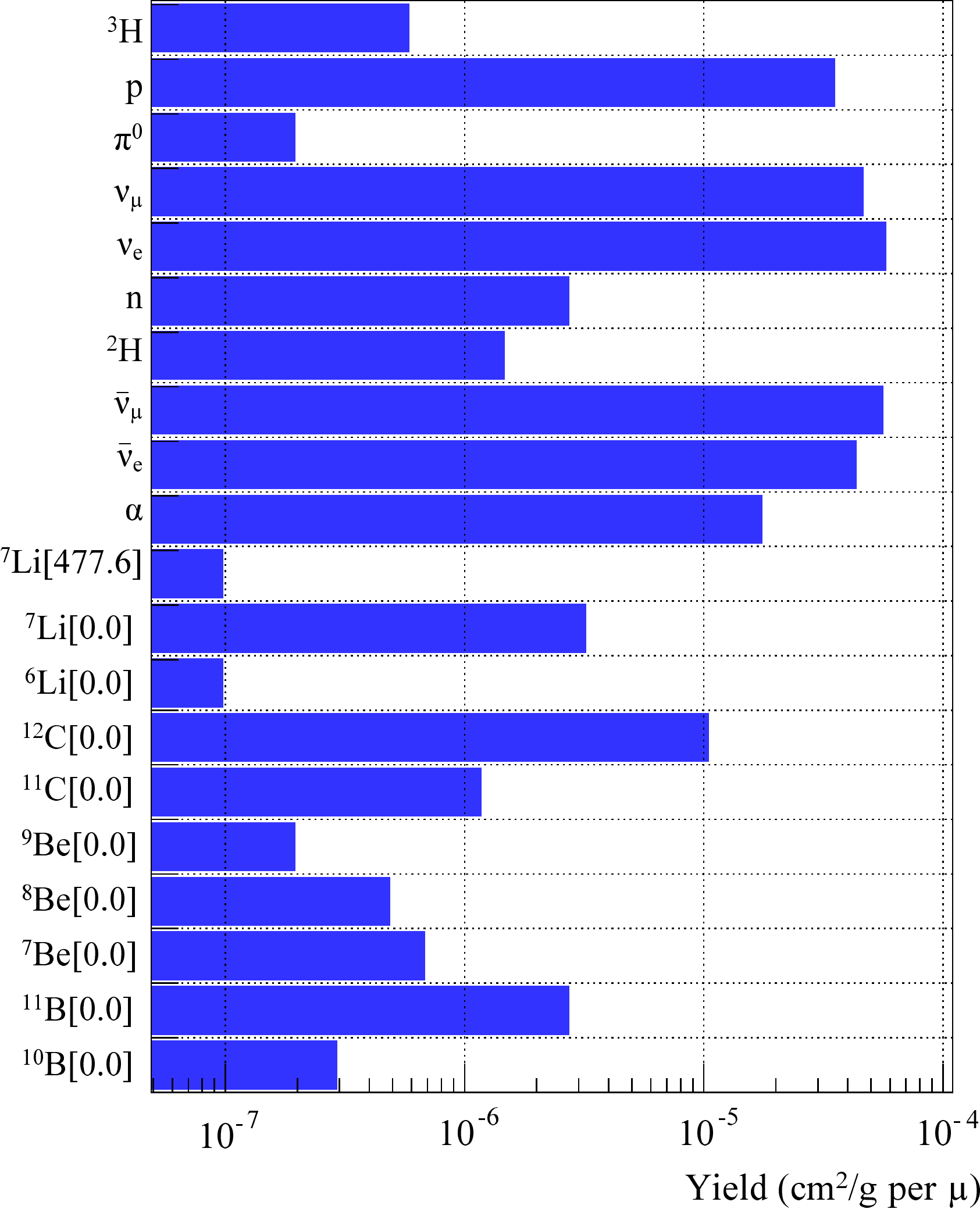}
\end{center}
        \caption{
                Cosmogenic isotope production yield due to 
                sea-level spectrum
                $\mu^{-}$s passing through 10 m 
                of EJ-254 plastic.
                $10^{4}$ events were simulated for this result. The number enclosed
                in brackets in the labels along the y-axis is the excitation energy of
                the isotope in units of keV. 
                $\nucl{9}{Li}$ and $\nucl{8}{He}$ were not observed.
        }
        \label{fig:cosmogenic_isotope_yield}
\end{figure}

In order to study this problem in more detail, a GEANT4~\cite{Agostinelli:2002hh} simulation of the EJ-254 plastic scintillator was conducted. Sea-level spectrum cosmic ray muons were incident on a $(10\times10\times10)~\mathrm{m^3}$ cube of scintillator. The isotope yield per muon event for this simulation run is tallied and shown in Fig.~\ref{fig:cosmogenic_isotope_yield}.

\begin{figure}
\includegraphics[width=1.0\linewidth]{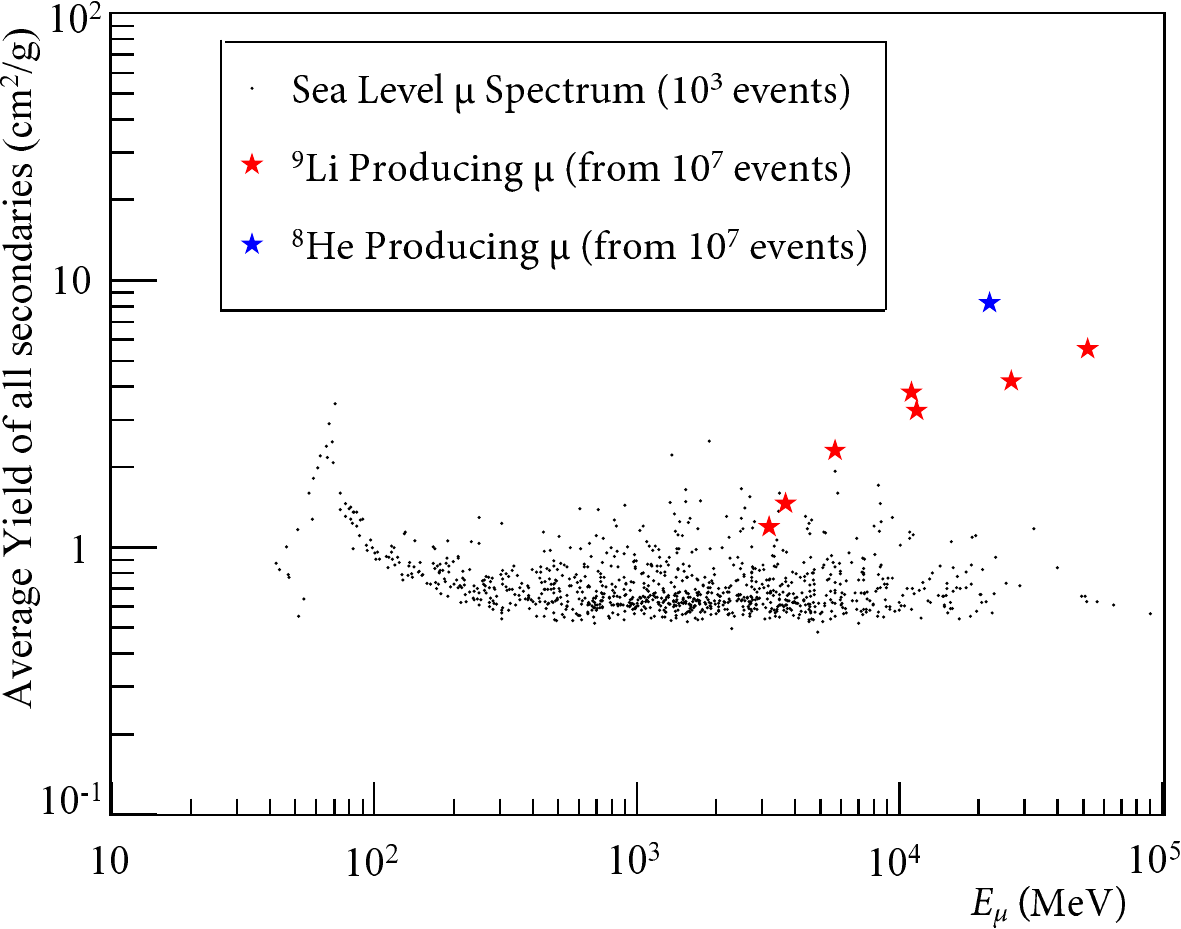}
\caption{Average yield per unit muon track length per unit medium density of all
non-photon secondary particles versus muon energy. The black points show the
result of a run of $10^3$ sea-level spectrum muons. Superimposed on the figure
are eight specific muon events that had produced a $\nucl{8}{He}$ or $\nucl{9}{Li}$
daughter depicted by the colored stars. These eight events were extracted from a
much larger ensemble with an increased statistics of $10^7$ events in order to
produce the rare events.
\label{fig:mu_backgrounds_He8_Li9}}
\end{figure}

Figure~\ref{fig:mu_backgrounds_He8_Li9} shows the average secondary particle
yield per unit muon track length per unit medium density for simulated cosmic
ray muons using a sea-level energy spectrum. The result implies an isotope yield
of $\sim 6.86 \times 10^{-10}$~cm$^2$/g for $\nucl{9}{Li}$ and \mbox{$\sim 9.79
\times 10^{-11}$~cm$^2$/g} for $\nucl{8}{He}$. The atmospheric muon rate
traversing the 13~cm cube is about 1/s (depending upon overhead shielding), and
so the rate of these events being produced in the mTC is estimated to be less
than 1~event per year. In addition, the general behavior of typical sea-level
spectrum muons shown by the black points involves a relatively constant
production of secondaries with respect to energies above a few hundred~MeV;
whereas a trend of increasing daughter production is clearly seen for those
producing $\nucl{9}{Li}$. This suggests that the $\nucl{9}{Li}$ isotope is most
likely produced in showering muon events at high energies, which can be easily
vetoed. Rejection of backgrounds associated with $\nucl{8}{He}$ will require
more statistics and further investigation.

Peripheral geometries of the detector and its in-situ environment pose a non-negligible contribution to the cosmogenic backgrounds and a more accurate study with these effects fully taken into account will need to be conducted in the future. Design and production of the shielding cave is currently underway and these background studies will be pursued in parallel as development continues.

Finally, with an mTC-type detector, it is straightforward to implement additional vetos to reject backgrounds if needed. This can be accomplished by installing scintillator paddles around the mTC and feeding additional trigger signals to the trigger/clock distribution board.

\subsection{Backgrounds at the NIST Reactor Location}

\begin{figure}
\includegraphics[width=1.0\linewidth]{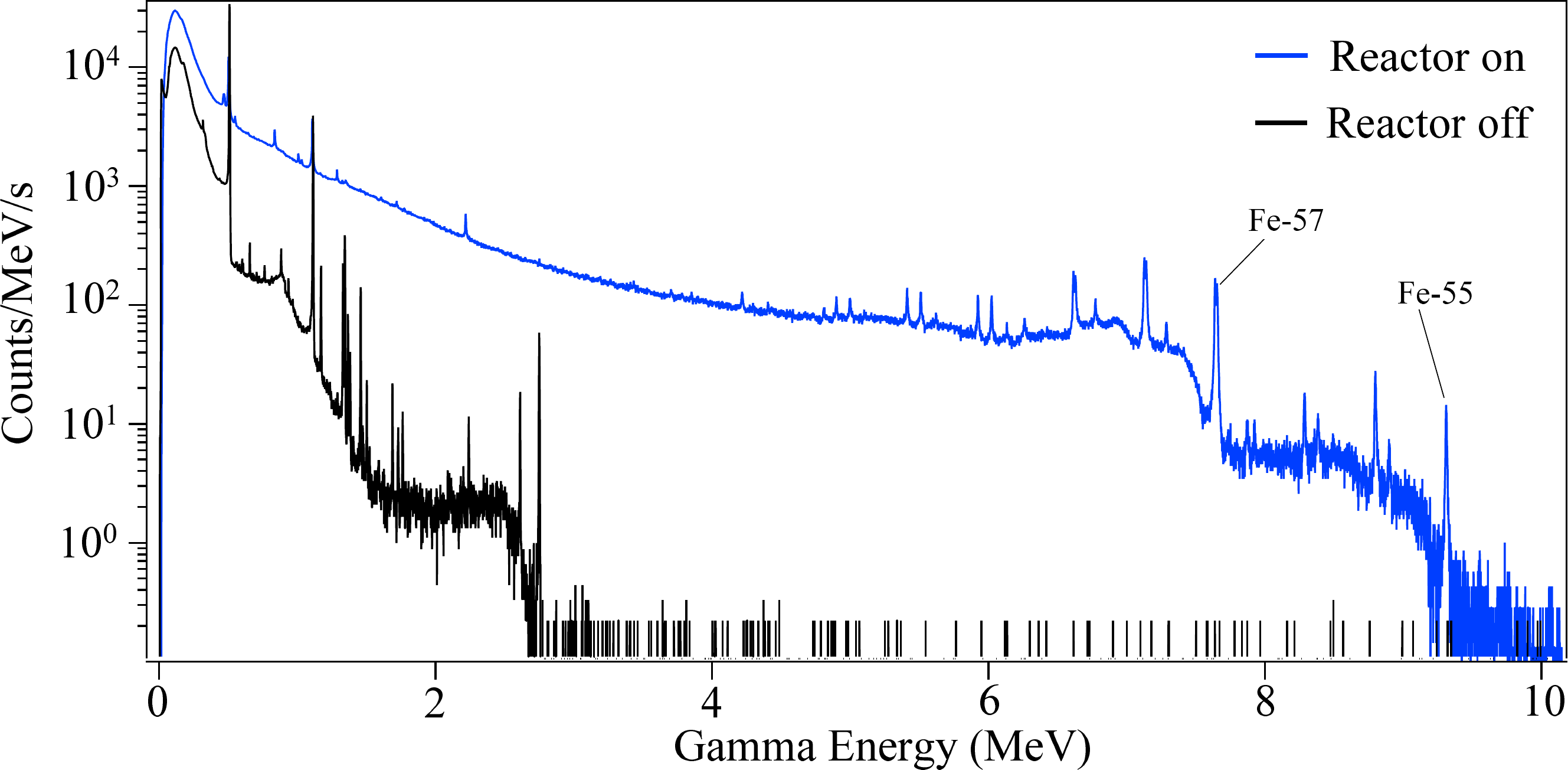}
\caption{ High Purity Germanium (HPGe) gamma-ray spectrometer response at the mTC location adjacent to the NIST reactor. HPGe spectrometer is 55~mm in length and 62.5~mm in diameter. The observed Fe lines are from neutron capture on surrounding shielding and structural materials.
\label{fig:NIST_Gammabkg_1}}
\end{figure}

Detailed background studies must be performed at a particular reactor site, since all venues differ and the backgrounds depend in detail upon local conditions. A group preparing for the PROSPECT experiment carried out a detailed background survey in the mTC location.\cite{Ashenfelter:2015tpm}

Figure~\ref{fig:NIST_Gammabkg_1} shows the gamma spectrum at the proposed mTC location without shielding.  The difference between the reactor on/off spectra is readily visible. The ``reactor on'' spectrum extends to relatively high energies due to prompt gammas from neutron capture thus posing additional challenges for shielding.

\begin{figure}
\includegraphics[width=0.7\linewidth]{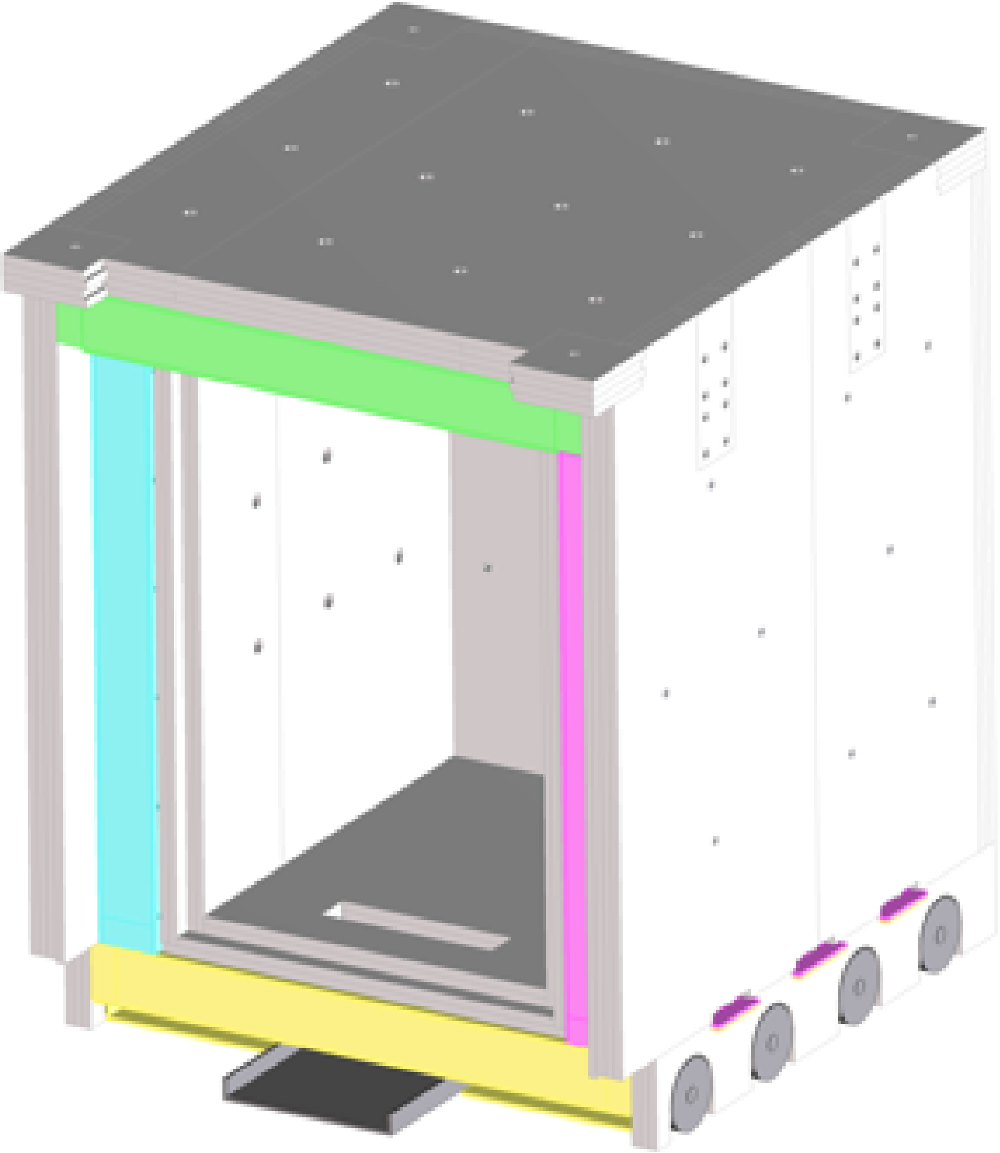}
\caption{CAD of the multi-layer shielding for the mTC.
\label{fig:CAD_shielding}}
\end{figure}

Deployment of mTC as an antineutrino detector at the NIST reactor, where the backgrounds are particularly high due to adjacent neutron scattering instruments, requires shielding from various background signals that could overwhelm or create false events within the scintillating volume (e.g., high-energy gammas, thermal neutrons, fast neutrons, cosmic ray induced muons and their decay products). In most other anticipated deployment locations we do not expect shielding to be critical. 

Towards that effort a multi-layered shielding cave was designed to encase the mTC and most of its associated electronics during the testing at the NIST reactor. The mTC detector will be deployed inside this shielding cave and together they will be placed against the face of the reactor biological shield. The shielding cave is comprised of six nested cubes, with the outermost dimensions yielding a footprint of 
\mbox{$\sim$ 1.8~m $\times$ 2~m $\times$ 2.7 m} and a total internal wall thickness of roughly 0.4 m. From exterior to interior, the layers are as follows (Fig.~\ref{fig:CAD_shielding}): 

\begin{enumerate} \itemsep1pt \parskip0pt \parsep0pt
 \item 10 cm of 5\% borated polyethylene sheet 
 \item 1 cm of A36 steel plate 
 \item 15 cm of steel shot and paraffin wax mixture 
 \item 1 cm of A36 steel plate 
 \item 10 cm of 5\% borated polyethylene sheet 
 \item Interior cavity for housing the mTC and associated electronics (dimensions \mbox{1~m $\times$ 1.2~m $\times$ 1.5~m})
\end{enumerate}

Borated polyethylene was chosen for its neutron absorbing properties while the layer of steel shot and wax acts as both a neutron and gamma absorber. All layers serve to attenuate the muon flux, albeit less efficiently. The overlapping construction removes potential line-of-sight and the hermetic design inhibits the penetration of thermal neutrons, which exhibit gas like properties.  

The shielding cave is 20 tons. The entire weight of the cave is supported on rails already laid into the floor at the NCNR, which allows the cave to be moved across different baselines.

The interior of the shielding cave will house the mTC, complete with its electronics rack containing high voltage power supplies and support electronics. The system requires less than 2 kW of 115 VAC power, and has its own uninterruptible power supply system. Accommodations for cooling of the electronics will be used, with access for cooling and electricity through a floor tray. During mTC operation access to the interior of the cave is expected to be infrequent.

We use GEANT4 to estimate effectiveness of the different shielding layers in attenuating potential backgrounds.  The modeled environment includes the shielding cave, a 0.5~m thick concrete roof above the area, and a concrete reactor bio-shield next to the shielding cave.
Three primary particles (muons, neutrons, and gammas) and two particle sources (atmospheric and reactor), including their relevant energy spectra and fluxes, were used.

Atmospheric gamma and muon energy spectra were calculated using the Cosmic-RaY shower Library (CRY),\cite{CRY}.
We use the spectrum and flux for atmospheric neutrons from Gordon {\it et al.}\cite{2004ITNS51:3427G} A Maxwell-Boltzmann distribution at 600 K with an integrated flux of 3-4 neutrons cm$^{-2}$ s$^{-1}$ was used to represent reactor neutrons outside the shielding cave.  The integrated flux was chosen to match energy-insensitive Bonner ball measurements taken {\it in situ}.  Because the neutron energy spectrum was not measured, a higher characteristic temperature (vs 298 K) allows the simulation to conservatively account for a portion of the reactor neutrons not being thermalized. The reactor gamma spectrum and flux ($\sim$200 cm$^{-2}$ s$^{-1}$ above 100 keV) comes from a measurement at NIST adjacent to the proposed mTC site with the reactor on.  

For the purposes of preliminary Monte Carlo work, atmospheric particles were assumed to follow a $\cos^{2} \theta$ angular distribution.  Reactor particles were assumed to be isotropic, although significant spatial variation coming from localized source has been measured, and if needed will be incorporated in later work.

Preliminary estimates of the resultant particle fluxes through the mTC volume with and without the cave present are shown in Table~\ref{table:mTC_cave}. These include secondaries produced within the shielding material itself. Measurements taken {\it in situ} useful for Monte Carlo validation are planned and will be reported in a future publication.

\begin{table}[h]
\caption{Particle fluxes through the mTC volume with and without the shielding cave present. The neutron flux is dominated by near-thermal neutrons hence the large attenuation factor.  Muons are incident on the mTC at a rate of less than 3.5~Hz, and are not significantly affected by the shielding cave.\label{table:mTC_cave}}
\begin{tabular}{ |l || c  c | c  c | c| }\hline
    	 	& normal 	& shielded 	& normal 			& shielded 		& attenuation 	\\
\cline{2-6} 	Type	& \multicolumn{2}{|c}{\#/mTC/s} & \multicolumn{2}{|c|}{\#/cm$^2$/s} 				&  	\%		\\ \hline\hline
Neutron 	& 3391 	& 0.082		& 4.0 						& $9.7 \times 10^{-5}$ 		& 99.9\% 	\\
Gamma 		& 169015 		& 325 		& $2.0 \times 10^{2}$			& $3.8 \times 10^{-1} $ 	& 99.8\% 	\\ \hline
\end{tabular}
\end{table}

The Monte Carlo model of the mTC using the shielded fluxes in Table \ref{table:mTC_cave} shows a signal to background (for uncorrelated events only) of roughly 1:1. These uncorrelated events are usually composed of two independent gammas entering the detector within our 12 $\mu$s time window, the first creating a false prompt signal and the second a false delayed signal. An order of magnitude less likely are uncorrelated backgrounds in which a neutron creates a false delayed signal instead of a gamma. These simulation results indicate that accidental coincidences from uncorrelated backgrounds will likely not be our dominant background source, and we are beginning to focus more on correlated secondaries originating from high-energy cosmogenic neutrons and muons.

\section{Electronics}
\label{sec:Electronics}

The mTC concept puts stringent requirements on the channel density, timing performance, synchronization, and power consumption of the detector. 
In order to fully utilize the spatial information provided by the pixelization of the 24 MCP-PMTs, all 1536 channels must be separately instrumented.
The readout for each pixel must preserve the $\mathcal{O} (100 \mathrm{ps})$ timing provided by the photodetector, and to avoid further timing degradation the timing of all detector channels must be synchronized to one another at a level significantly below the transit time spread of the MCP-PMT.
An online trigger system is required to isolate physics interactions of interest from backgrounds. 

\subsection{Front-end Electronics}

The core of the front-end electronics functionality is provided by the IRS,
\footnote{The Ice Radio Sampler (IRS) is a descendant of the Buffered LABRADOR series of ASICs,~\cite{Ruckman:2008fx} and was originally designed
for performing radio searches for ultra-high energy neutrinos using antennas embedded in Antarctic ice.}
a family of application specific integrated circuits (ASICs) developed at the University of Hawaii.
The IRS has been used in a variety of projects that require fast sampling and deep buffering.\cite{Dey:2014fda,Bechtol:2011tr,Allison:2011wk} 

\begin{figure*}[htbp!]
\includegraphics[width=1.0\linewidth]{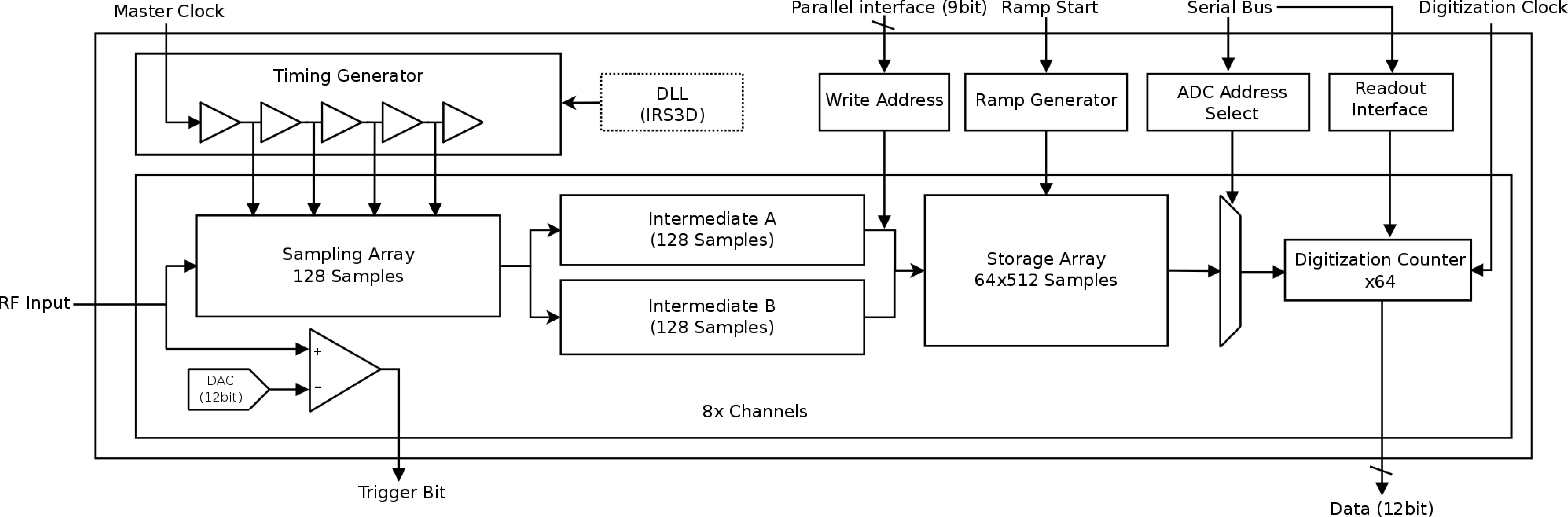}
\caption{Block diagram of the IRS ASIC architecture.  Eight channels of analog input are received by a set of eight sampling arrays, with sampling timing based on a common timing generator, driven by an external clock.  This timing generator also determines timing of transfers from the sampling arrays to intermediate and storage arrays.  The target location for the transfer from the intermediate to storage array is controlled by the user with a 9-bit parallel bus.  A separate pin is used to start an internal voltage ramp, used to digitize 64-samples of the storage array for all eight channels in parallel.  Selection of the storage address to digitize is controlled through a serial interface.  A clock for the Wilkinson digitization process is generated internally (IRS3B) or provided externally (IRS3D).  Once data is digitized, the channel and sample to readout are controlled by a second independent serial interface.  Digitized data is available on a parallel 12-bit bus.  A number of DACs and internal timing parameters are controlled by a third serial register interface.}
\label{fig:IRS_diagram}
\end{figure*}

\begin{table}
[h]
  \caption{Operating parameters for the IRS family of ASICs, and nominal ASIC operating conditions for the mTC. Full performance parameters will be reported in a future publication.\label{table:IRS_parameters}}
  \begin{tabular}{| l || c | c | c |}
  \hline
  Parameter & IRS Range & mTC Setting \\ 
  \hline
  \hline
  Channels & \multicolumn{2}{| c |}{8} \\
  Sampling cells & \multicolumn{2}{| c |}{128} \\
  Storage depth & \multicolumn{2}{| c |}{32,768} \\
  Analog bandwidth & \multicolumn{2}{| c |}{$ > 300$~MHz} \\
  Digitization & \multicolumn{2}{| c |}{on-chip Wilkinson} \\
  Quantization & \multicolumn{2}{| c |}{12(9)-bits logged(effective)} \\
  Dynamic range & \multicolumn{2}{| c |}{$\sim 2~\mathrm{V}$} \\
  Typical noise & \multicolumn{2}{| c |}{$\sim 1~\mathrm{mV}_{\mathrm{RMS}}$} \\
  \hline
  Sampling rate & \multicolumn{1}{| c |}{1--4~GSa/s} & 2.73~GSa/s \\
  Master clock & \multicolumn{1}{| c |}{8--31~MHz} & 21.3~MHz \\
  Buffer time & \multicolumn{1}{| c |}{$(8-32)\mu\mathrm{s}$} & $12.0~\mu{\mathrm{s}}$ \\
  Conversion time & \multicolumn{1}{| c |}{$>2~\mu\mathrm{s}$} & $6.2~\mu\mathrm{s}$ \\
  \hline
  \end{tabular}
\end{table}

\begin{figure} [!htbp]
  \includegraphics[width=1.0\linewidth]{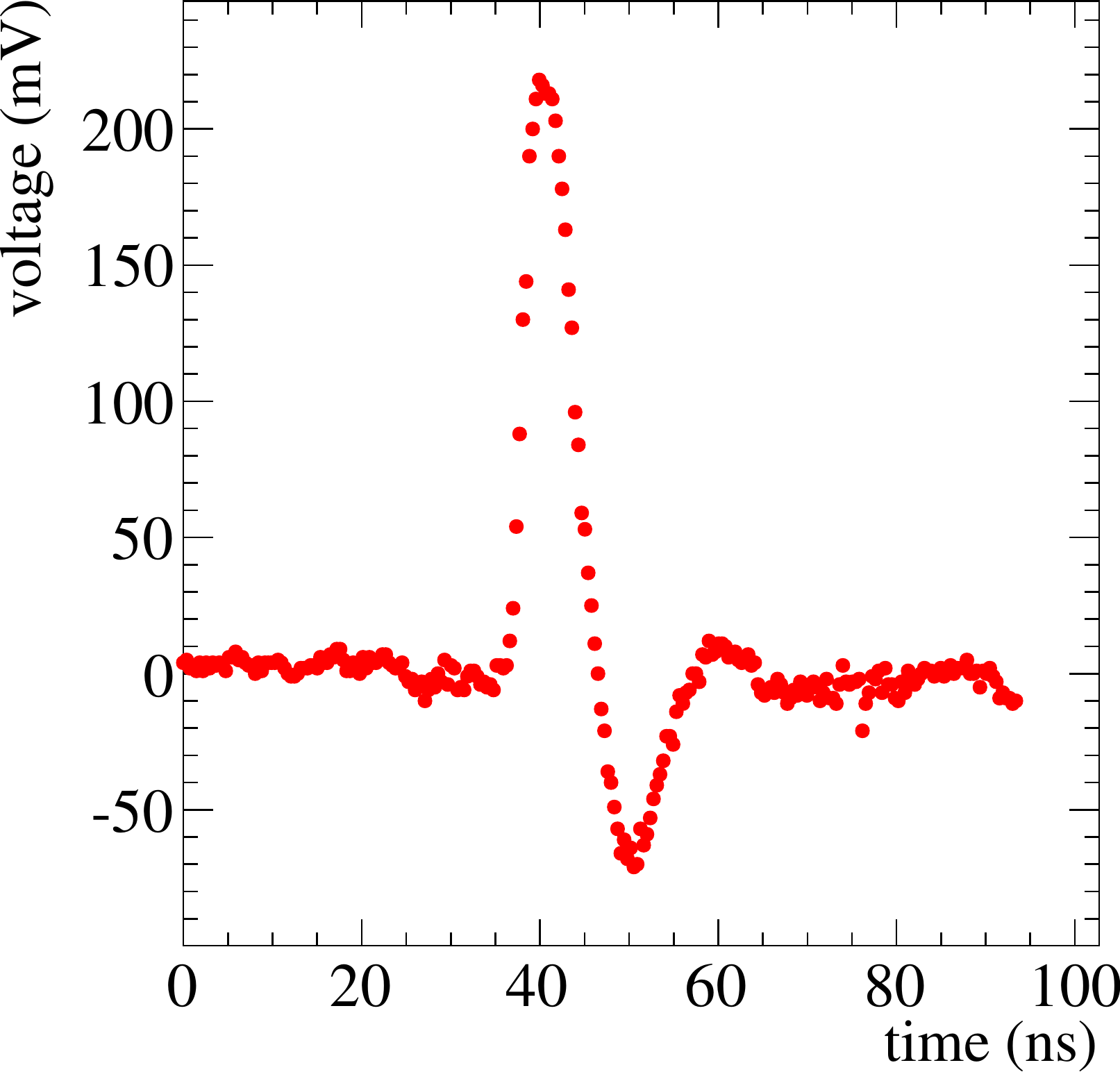}
\caption{Example MCP pulse digitized with the IRS3B.  Times for each point are calculated based on a nominal sampling rate of 2.73~GSa/s.  Voltages are calculated based on a nominal conversion factor of 0.6~mV/ADC count, and represent the signal after passing through an external amplifier.
\label{fig:irs3bWaveform}}
\end{figure}

The IRS ASIC architecture is shown schematically in Fig.~\ref{fig:IRS_diagram}, and a list of operating parameters can be found in Table~\ref{table:IRS_parameters}.
The ASIC has 8 analog input channels, each with a sampling stage, intermediate and deep storage stages, on-chip digitization, and per-channel threshold triggers.  
The sampling stage is a multi-GSa/s switched capacitor array (SCA) waveform sampler, similar to other ASICs,~\cite{Delagnes200621,Ritt2010486,Oberla:2013mra} in which a sampling clock propagates down a delay line, with subsequent delay stages utilized to create short, GHz-scale timing intervals to sample the input signal onto capacitors.   
Unlike other SCA waveform samplers, this sampling array is connected to deeper buffers to allow for higher trigger latencies and larger time records per-event.  
Buffer amplifiers are used to drive the stored voltages from the sampling array into a deep sampling array consisting of 32,768 storage capacitors per channel.  
This transfer occurs via an intermediate storage array to accommodate the settling time of the buffer amplifiers.
Signals required to coordinate the intermediate transfers are provided by an internal timing generator, and the final location of the samples in the deep storage array is provided by a parallel address bus that is driven by the user, allowing for flexible and user-defined memory management schemes.
The IRS includes 12-bit Wilkinson ADCs, which digitize 64-sample blocks of the storage array for all 8 channels in parallel.
Readout of the digitized data is done one sample at a time through a 12-bit parallel bus.  
Selection of the channel and sample number is provided by the user via a serial interface.  A typical digitized MCP pulse in the mTC system is shown in Fig.~\ref{fig:irs3bWaveform}.

Analog inputs for each channel are also monitored by a comparator, with the digital trigger bits available to the user.
These bits can be used to monitor which sections of analog memory have signals above a user-defined threshold, allowing the user to select only those windows which have signals of interest to be read out.
Further details on the IRS ASICs and their performance will be presented in an upcoming publication.

\begin{figure} [htbp!]
\begin{overpic}[width=1.0\linewidth]{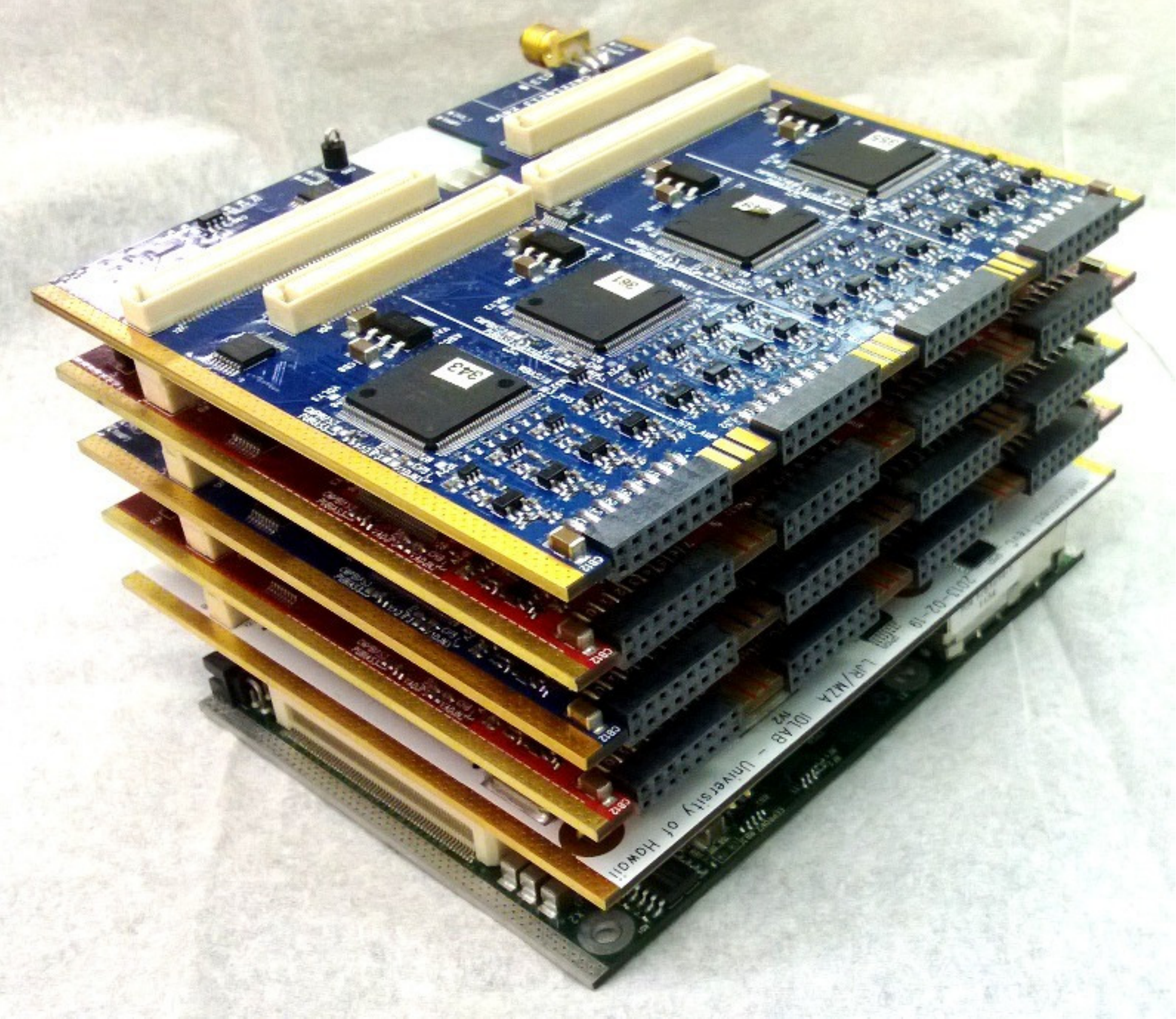}
\linethickness{3.0pt} 
\put(8,17){\color{black}6 cm}
\put(5,5){\color{black}\line(0,1){35}}
\end{overpic} 
\caption{A photograph of one of the twelve board stack assemblies. The 128 MCP signals are input via connectors shown. Each board stack instruments a pair of MCPs and is enclosed in an aluminum cage surrounded by chiller plates.}
\label{fig:board_stack_photo}
\end{figure}

One front-end electronics module, or ``board stack," shown in Fig.~\ref{fig:board_stack_photo}, includes 16 IRS ASICs (128 total input channels from 2 MCPs).
Each analog input is amplified by an RF amplifier before arriving at the IRS ASIC.
The initial version of the mTC was developed with the IRS3B ASIC, and an upgrade is in progress to move to new board stacks using the IRS3D, a new revision of the ASIC that includes improvements to reduce noise and improve timing stability.   
In both versions of the front-end electronics, a single FPGA (Xilinx XC6SLX150T) on each board stack provides all control signals necessary to operate and readout the ASICs and other auxiliary devices.
The FPGA interfaces to the back-end data acquisition system for register control and data transmission via fiberoptic cable.  
To coordinate timing between the 12 modules of the mTC, each board stack accepts a central distributed clock via RJ45 connector.
Another RJ45 connector is wired to the FPGA JTAG interfaces, allowing remote programming of the FPGA firmware in-situ.

\subsection{Clock Distribution and Triggering}

All 12 board stacks communicate with a custom PCB, designated Clock and JTAG In PCI (CAJIPCI), over differential pairs of CAT7a cable.
The CAJIPCI provides a low jitter ($\sigma_t < 2~\mathrm{ps}$) clock to the front-end modules.
Front-end board stacks provide a module-level trigger to the CAJIPCI over another differential pair on the same cable, and the CAJIPCI responds with a system level trigger over a third pair.
The final differential pair can be used to perform flow control and limit trigger rates to the front-end electronics.

Three separate trigger levels are defined for the experiment.  
The lowest level trigger is a level 0 (L0) trigger, defined as a single channel trigger bit from an IRS ASIC.  
Thresholds for these triggers can be set via adjustment of an on-chip DAC. 
The 128 L0 triggers on a board stack are monitored by the FPGA.  When the number of coincident triggers falls between two user programmable thresholds, a level 1 (L1) trigger is issued and sent to the CAJIPCI.
The CAJIPCI, in turn, monitors L1 triggers from the 12 front-end board stacks, and issues a level 2 (L2) trigger to the front-end modules under user-defined conditions.

\begin{figure} [!htbp]
  \includegraphics[width=1.0\linewidth]{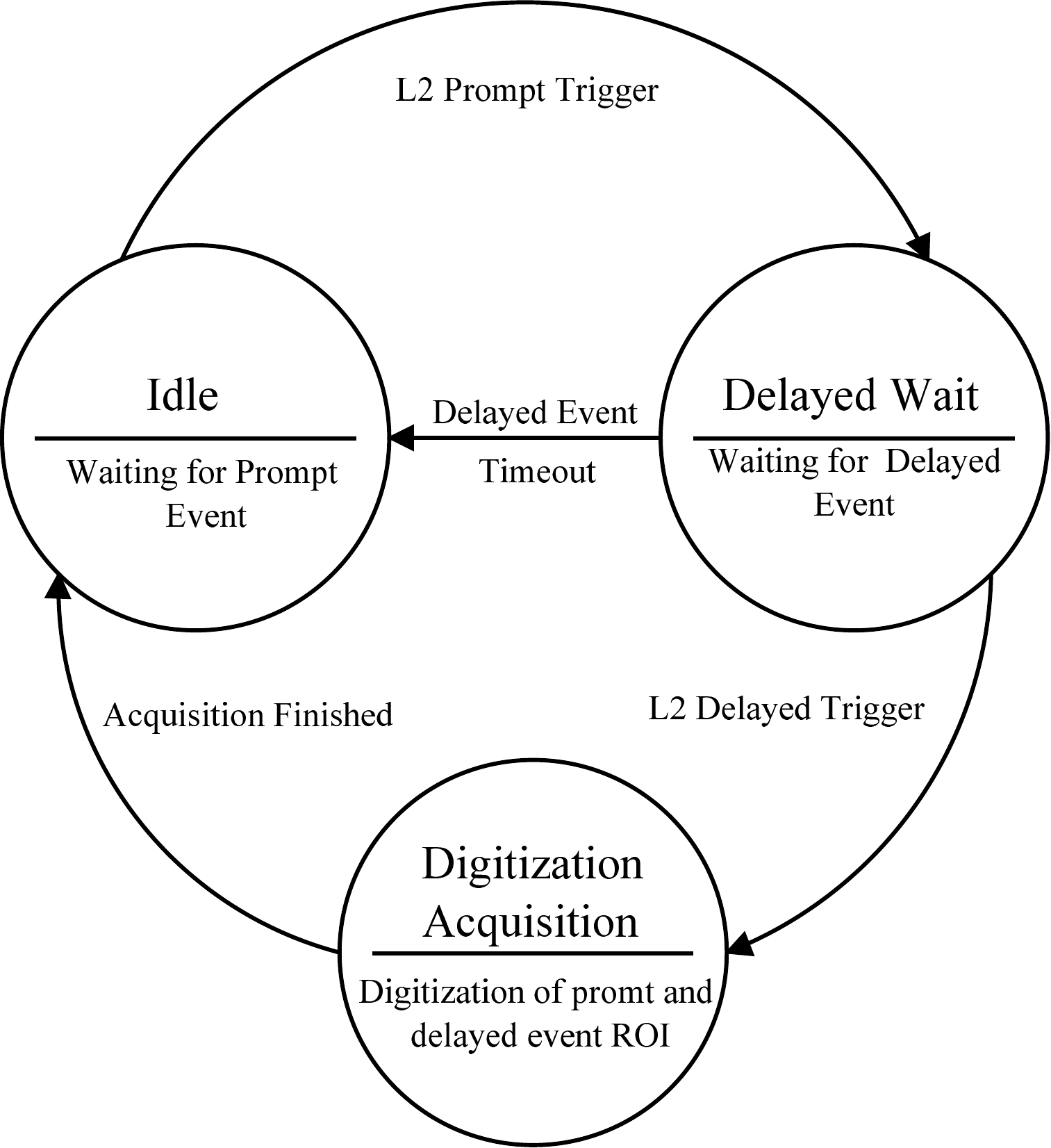}
\caption{State machine diagram of the neutrino trigger.
\label{fig:state_machine}}
\end{figure}

A basic L2 trigger can be calculated based on the number of coincident L1 triggers.
This basic trigger is appropriate for signals that fall mainly in a narrow time window (e.g., neutrons, gammas, and cosmic ray muons).  A neutrino L2 trigger must monitor for both a prompt and delayed signal, so it includes an initial "arming" period when it detects a prompt signal, and a second stage to issue a trigger upon receipt of a delayed signal.
This logic is shown in Fig.~\ref{fig:state_machine}.
The timeout for the delayed trigger is typically set to $\sim12~\mu\mathrm{s}$, the length of the IRS storage array.
Longer times between prompt and delayed signals are possible based on the analog memory management scheme used for the IRS, and this may be explored in future upgrades.

\subsection{Data Acquisition and Software}

Upon receipt of an L2 trigger, data from the front-end modules is digitized, readout by the front-end FPGA, and sent over fiberoptic cables using a gigabit Ethernet (1 GbE) interface.  This data is received by commercial PCIe Ethernet cards, running on a rack-mount server, which can be operated directly or via a network connection. Operations from powering up to collecting data may be performed remotely over this network, with an aim for full remote control. 

Data acquisition is implemented primarily in C++, with a Python API to the system that can be used for configuration as well as monitoring and slow control of the detector.
Several algorithms and programs have been developed to perform automated startup, initialization, and real-time fine tuning of the  electronics.
Before physics data acquisition begins, threshold scans and pedestals are collected for each channel and stored in files for each board stack.
Timing parameters for each chip are taken, adjusted, and stored to ensure optimum calibration.
Once this procedure is completed, the files can be used for repeated data runs.
For more detail on the required electronics calibrations, see Section \ref{sec:Calibration}.

Once the detector is initialized and data is taken, the data can be observed and analyzed with tools developed by the collaboration. Several event viewers have been developed and analytical methods for analysis and event reconstruction are being employed using packages such as MATLAB, C++, and ROOT.

\subsection{Support Systems}

Power is supplied by high voltage (HV) and low voltage (LV) units from modular power supplies.
These are mounted in the rack underneath the mTC's aluminum enclosure and are operated remotely. 
The front-end power consumption is approximately 330 W, so cooling of the electronics is crucial.
Commercially available hard-drive chiller plates are used and mounted on the
electronics card cages with deionized water as a coolant at
a total flow rate of $\sim 2$~GPM.  
When the detector is operating, the temperature
of the ASICs is monitored during operation to ensure the safety and stability of the electronics, and typically is stable in the range $\sim$~30--35~\textcelsius, depending on ASIC position.

\section{Calibration}
\label{sec:Calibration}

A number of calibrations are required to operate the detector and remove systematic biases, including electronics effects (both amplitude and timing), MCP efficiency and gain, and calibration with physics processes. 
We describe each in more detail here.

\subsection{Electronics Calibrations}

The architecture of the IRS ASIC utilizes individual capacitors and comparators for each of the 32,768 storage cells of an input channel.
Variations in the fabrication process create sample-to-sample differences in threshold voltages for the comparators, resulting in a fixed-pattern voltage structure that must be removed from digitized waveforms.
These are known as ``pedestals" and are evaluated by collecting events with no signal input.
This may be done, for example, with the MCP high voltage turned off, or with the high voltage on but using software triggers that are uncorrelated with any signal inputs.
Pedestals are typically collected at the beginning of a run period.
Over $50\times10^{6}$ pedestals are required to run all channels of the detector at their full sampling depth.
An example waveform following pedestal subtraction can be seen in Fig.~\ref{fig:irs3bWaveform}.

Further feature extraction is performed on pedestal subtracted data, including estimates of pulse height and pulse times.  Pulse timing is estimated using an offline, software-based, constant-fraction discrimination method, with time defined by the crossing of the signal over a set percentage of the pulse height, typically around 50\%.  Linear interpolation is used to determine this time with much higher granularity than the 370~ps spacing of the individual samples.
To achieve the best possible timing resolution, further calibrations are necessary to remove ASIC fabrication effects.
The delay line used to generate the fast sampling signals within the ASIC is a current-starved inverter chain.
As with the storage cell comparators, process variations cause threshold variations in these inverters, leading to non-uniform timing distributions from sample-to-sample.
This manifests as a fixed-pattern timing structure that is unique to each ASIC, which we refer to as the ``fine timing calibration.''
A total of 128 timing offsets must be calculated for each ASIC, one for each stage of the delay line.
Typical spreads in timing values are 10-15\% of the nominal sampling delay.
For example, in our standard operating mode with the IRS running at 2.7~GSa/s, the mean timing delay is 370~ps, with a spread of roughly $\sigma_t \approx 13-55~\mathrm{ps}$.

To perform this timing calibration we inject MCP-like pulses into the electronics at known delays relative to the sampling clock.
By stepping the delay of these pulses in fine increments (as low as 15~ps) we can calculate a pulse time in units of sampling cells and cross reference it against the known delay, allowing us to map out the fine time structure within the ASIC.
A total of 24,576 of these timing values (128 sample delays ~$\times$~ 16 ASICs ~$\times$~ 12 board stacks) are stored for the full detector.

Following these fine timing calibrations, we must then align all channels of the mTC to a common time reference.
Although all channels sample synchronously based on the distributed master clock, various skews are introduced throughout the system from the ASIC structure, PCB routing delays, cable lengths, etc.
To characterize these delays, we use a laser system, described below, to inject signal at fixed times into each face of the detector.  
By aligning timing of pulses for all channels relative to one another, we can measure these skews for each of the 1536 channels and remove them for any subsequent analysis.

\subsection{Laser Sources}

A precision timing laser system (Advanced Laser Systems EIG1000D) can inject signal into any of the 6 faces of the mTC through a `needle' fiber connector installed in the space between the MCPs.  
Variable neutral density filters can be inserted between the laser diode output and the input to the fiberoptic connections that inject into the mTC.  This allows studies at adjustable light levels, from single PE and up.
Stepper motors can be used to move optical elements and select the injection point of the laser, or adjust the laser attenuation.
The laser controller is triggered by the timing distribution board, allowing optical pulses to be injected at adjustable times relative to the master sampling clock.
These features allow the laser to be used as an automated in-situ calibration or validation source.

\subsection{MCP Gains}

All 24 MCPs operate on independent high voltage channels, allowing selection of gain independently for each tube.
A specification from the manufacturer is provided for each MCP unit with HV settings at $10^5$ gain.
We have further measured gain curves by observing the common-last-dynode of each MCP for laser and cosmic ray muon signals.
Specific HV settings vary by operating mode, as the expected number of PE detected covers a very broad range from tens of thousands of PE (e.g., for cosmic ray muons fully traversing the detector), to under 100 PE (for the delayed neutron capture from IBD).

Since gain can vary considerably across the pixels of an MCP, we must further calibrate the gains of each individual channel.
This is typically done by measuring single PE signals across the detector, either injected using the laser or by observing MCP dark pulses, and normalizing their mean amplitudes to one another.
This further provides a conversion factor from digitized counts to an estimated number of PE for each recorded pulse.
Quantum efficiency can be similarly calibrated on a pixel-by-pixel basis, using the calibration laser running in a mode where we collect primarily single PE pulses.

\begin{figure}[htbp!]
  \includegraphics[width=1.0\linewidth]{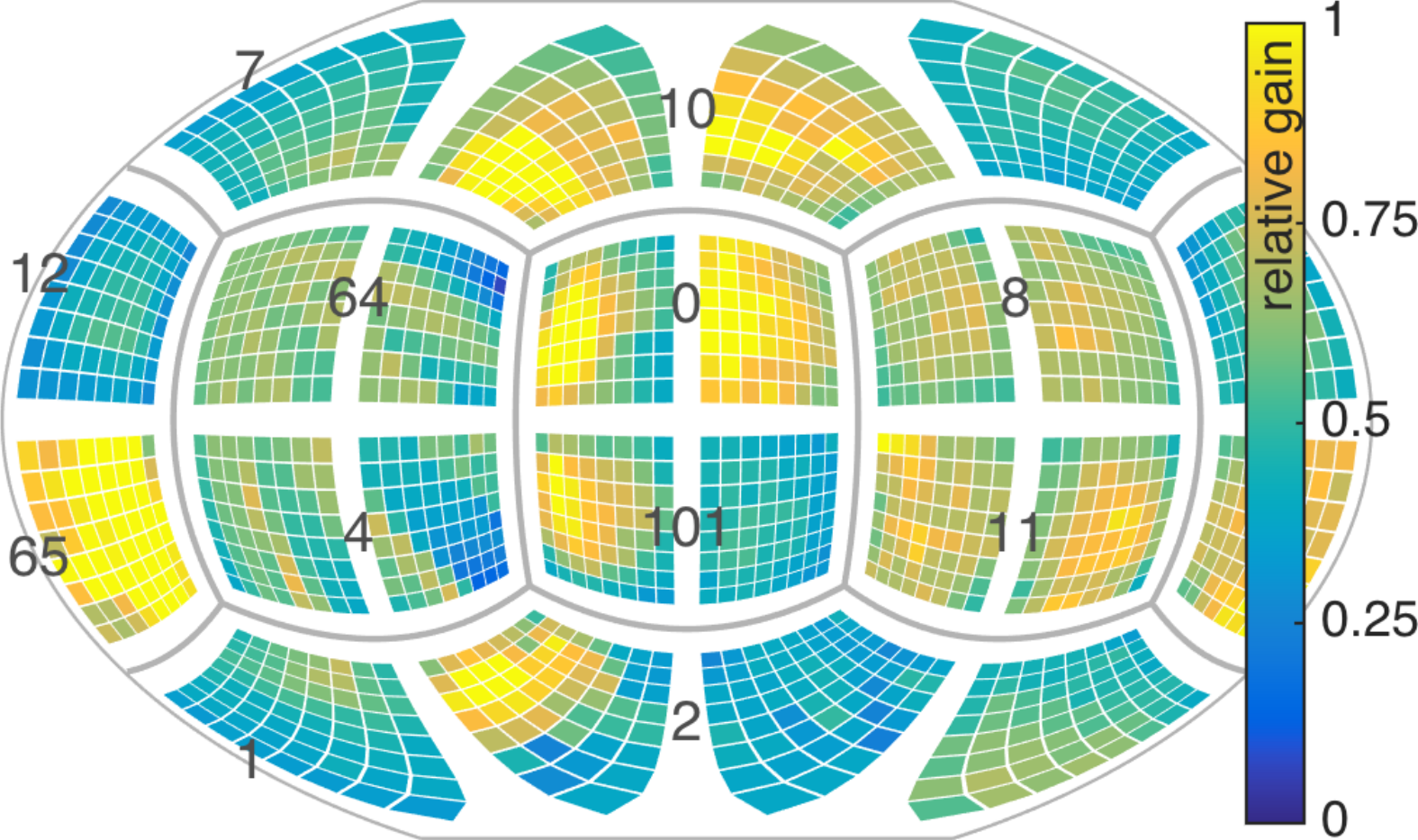}
\caption{
Display of a preliminary relative gain map obtained from laser data.  This includes electronics effects, such as variation in amplifier gain.
\label{fig:gain_map}}
\end{figure}

An example of a gain map obtained from laser data is shown in Fig.~\ref{fig:gain_map}.

\subsection{Cosmic Muons}

Cosmic ray muons provide one method for validating the calibrated performance of the detector, as they have a known energy deposition and time-stable rate. The fairly stable flux at sea level ($\sim 1$~Hz through the detector) has a mean energy of about 2 GeV and is peaked near the vertical but falls off towards the horizon gently with a $\cos^2$ of zenith angle. In typical running conditions one would expect some variation due to the local overburden.
The minimum ionizing energy loss rate for polyvinyltoluene~\cite{Agashe:2014kda} (the plastic in the mTC's scintillator)
is 1.956~MeV\;cm$^2$/g with density 1.02 g/cm$^3$, so the net (mean) energy loss rate in the cube should be about $dE/dx = 2.3$ MeV/cm.

\begin{figure*}[htbp!]
  \includegraphics[width=1.0\linewidth]{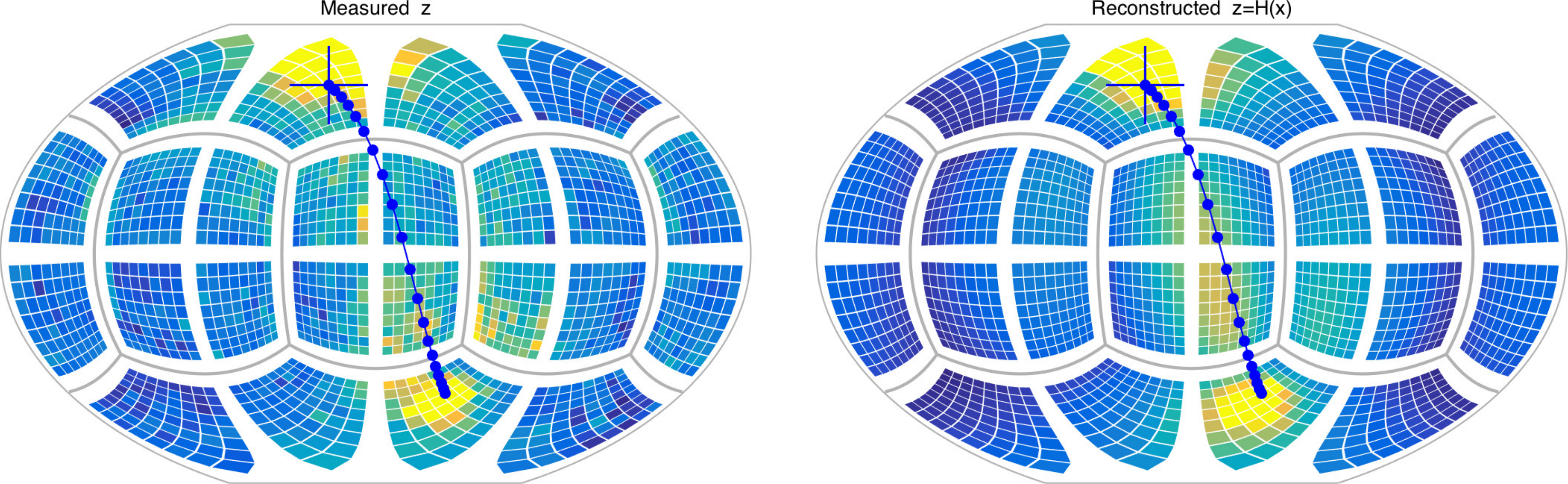}
\caption{
Event display for a muon measured in mTC (left) and the expected light distribution for the best fit reconstructed path of the muon (right).
\label{fig:cosmic_muon_rec2}}
\end{figure*}

\begin{figure}[htbp!]
  \includegraphics[width=1.0\linewidth]{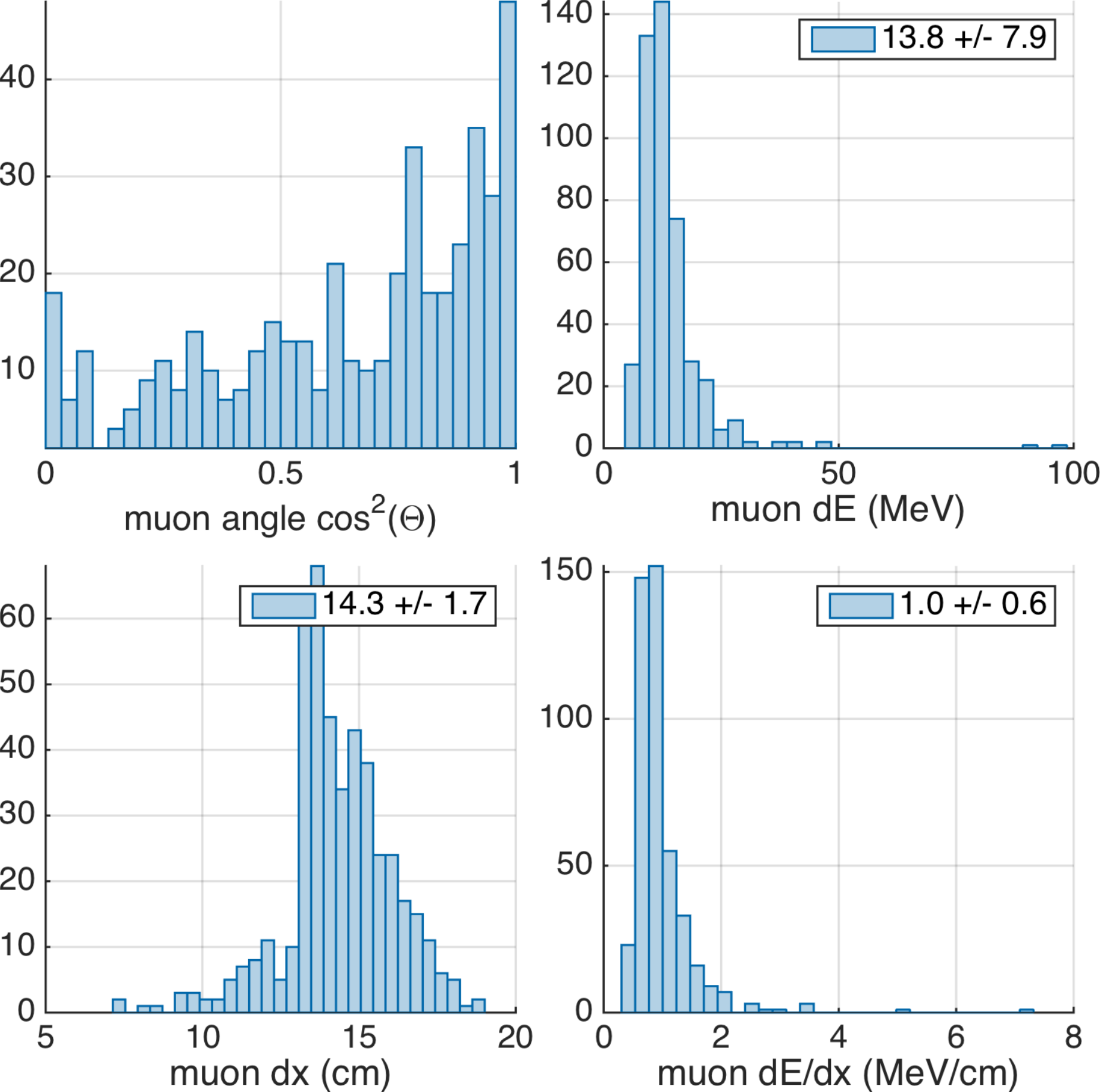}
\caption{
Preliminary distributions of reconstructed cosmic ray muon parameters using data collected with the mTC, showing reconstructed incident angle (top left), energy deposition (top right), muon track length within the mTC (bottom left), and muon energy deposition per unit track length (bottom right).  These distributions reflect a 500 event data sample.  Results are expected to improve as calibrations continue.
\label{fig:muon_parameters}}
\end{figure}

To acquire muon data, a low gain is set on the MCPs to avoid saturation, and the trigger levels are changed accordingly. 
The previously determined electronics and gain calibrations are used to analyze the resulting data, and muon tracks can be fitted through the detector, as shown in Fig.~\ref{fig:cosmic_muon_rec2}.  An example of reconstructed muon parameters for a preliminary data set is shown in Fig.~\ref{fig:muon_parameters}.

\section{Reconstruction}
\label{sec:Reconstruction}

The fast timing of the mTC's electronics, coupled with the excellent spatial resolution of the MCP channels, allows for high-quality reconstruction of subatomic events. Reconstruction is generally subdivided into two categories: unconstrained and constrained. {\it Unconstrained reconstruction} techniques like simple back-projection make possible the recovery in space and time of any arbitrarily-distributed pattern of energy, while {\it constrained reconstruction} techniques -- the simplest being a single point-source fit -- allow for the exploitation of {\it a priori} knowledge about the event, and provide more accurate reconstructions as long as correct assumptions are applied.\cite{mTC_neutron}

In general, the likelihood of observing a single photo-electron (PE) $z$ from a single point-source $\theta$ is
\begin{equation}
\label{eqnEst6}
p\left(z|\theta\right) = \Lambda_t P_\Omega P_\gamma P_T Q
\end{equation}

\noindent where $\Lambda_t$ is the temporal likelihood, $P_\Omega$ is the solid angle probability, $P_\gamma$ is the un-attenuated energy probability, $P_\mathrm{T}$ is the transmission (or non-reflecting) probability, and $Q$ is the PMT quantum efficiency. Equation \ref{eqnEst6} then forms the basis of our likelihood function, defining the likelihood of point-source $\theta$ given measurements $z$:
\begin{equation}
\label{eqnEst5}
p(\theta |z) = \prod_{j}^{}  p\left(z_j|\theta\right)p\left(\theta\right)
\end{equation}

\noindent where the likelihood $p(z_j|\theta)$ of measurement $j$ with prior $p\left(\theta\right)$ is simply an evaluation of the measurement space created by $\theta$ at $z_j$. Equation \ref{eqnEst5} extends to multiple point sources as well:
\begin{equation}
\label{eqnEst5i}
p(\theta |z) = \prod_{j}^{} \sum_{i}^{} w_i p\left(z_j|\theta_i\right)p\left(\theta_i\right)
\end{equation}

For point source $i$, the likelihood $p(z_j|\theta_i)$ of measurement $j$ given source $i$ with weight $w_i$ and prior $p\left(\theta_i\right)$ is simply an evaluation of the measurement space created by $\theta_i$ at $z_j$. This measurement space is defined by a point source position $P_\theta$ at time $t_\theta$, and is a function of several detector and scintillator characteristics including:

\begin{itemize}  \itemsep1pt \parskip0pt \parsep0pt
\item Scintillation spectrum, yield and decay constant(s)
\item Cherenkov spectrum
\item Quenching factors for heavy particles
\item Scintillator attenuation length
\item Re-emission efficiency of attenuated photons
\item Refraction indices of the scintillator and PMT glass
\item PMT QE
\item Time and energy calibrations
\end{itemize}

Equation \ref{eqnEst5i} forms the basis for a variety of parameter estimators in the mTC. Any number of complex shapes (i.e. muon tracks, neutron scatters, a complete antineutrino event) can be built up by using a collection of these simple point sources.

\subsection{Candidate Cuts}

\begin{figure}[htbp!]
\includegraphics[width=1.0\linewidth]{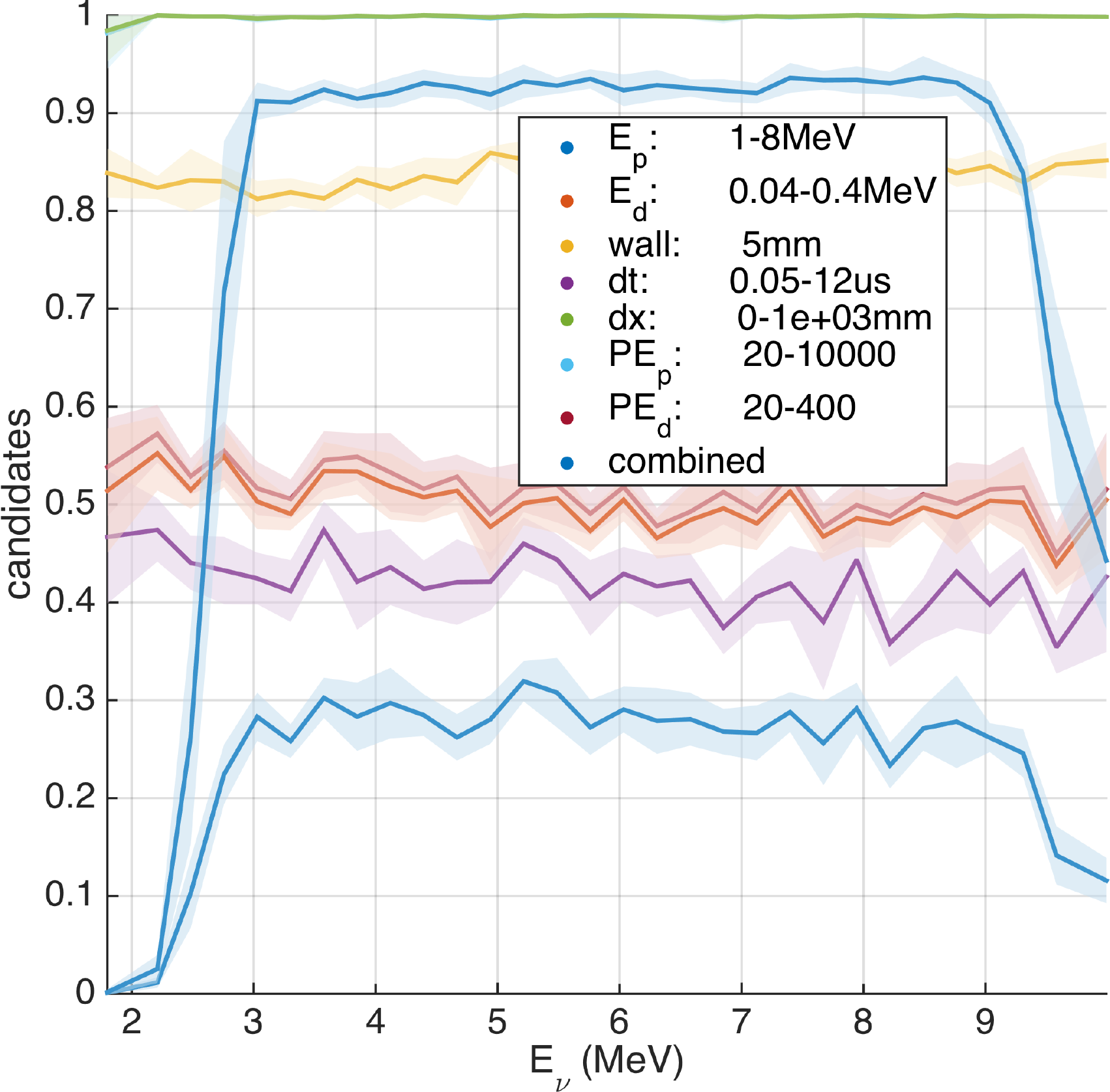}
\caption{Simulated mTC \nuebar~candidate efficiency vs.  \nuebar~energy.}
\label{fig:glennf4v}
\end{figure}

Measured events in the mTC pass through several candidate cuts before they are considered as possible \nuebar~candidates. These cuts, and their related candidate efficiencies are shown in Fig.~\ref{fig:glennf4v}. The cuts are implemented to both improve the quality of the \nuebar~events and also to reduce the likelihood of backgrounds entering into the \nuebar~candidate dataset.

The 5 mm edge cuts reject events with either prompt or delayed vertices \textless 5 mm from the wall. Since the mTC is a single volume detector, the MCPs are directly adjacent to the scintillation volume. Points which are fit too close to the wall tend to suffer from low reconstruction quality, and this cut serves to reject these potentially unreliable fits. Another reason for the edge cut is to reduce the likelihood of a positron from leaving the detector volume, which could result in severe underestimation of the \nuebar~energy. This edge cut reduces the detector fiducial volume by 20\%, from 2.2 to 1.7 liters.

There are time cuts on the prompt-delayed $dt$ as well; these are a 50 ns floor and a 12 $\mu s$ ceiling (hardware imposed). The floor is designed to prevent late prompt PE's from entering the delayed signal dataset. 

We impose energy and PE cuts as well. For the prompt signal we accept energies of 1-8 MeV and PE counts between 20 and 10,000 PEs. The delayed signal has much stricter energy cuts, as it has a more consistent energy output; we accept delayed candidates with between 20-400 PEs and 40-400 keV. 

\nuebar~candidates must meet all these requirements in order to be accepted into the \nuebar~candidate pool. In the mTC we find about 30\% \nuebar~candidate efficiency at 3-4 MeV. The dominant source of efficiency loss is neutrons leaving the detector volume, which happens 45\% of the time, and from neutrons leaving the 12 $\mu$s time window, which happens 30\% of the time. These two causes alone reduce the mTC \nuebar~candidate efficiency to \textless 40\%; the other cuts only have minor effect.

\subsection{Performance}

\begin{figure}[htbp!]
\includegraphics[width=1.0\linewidth]{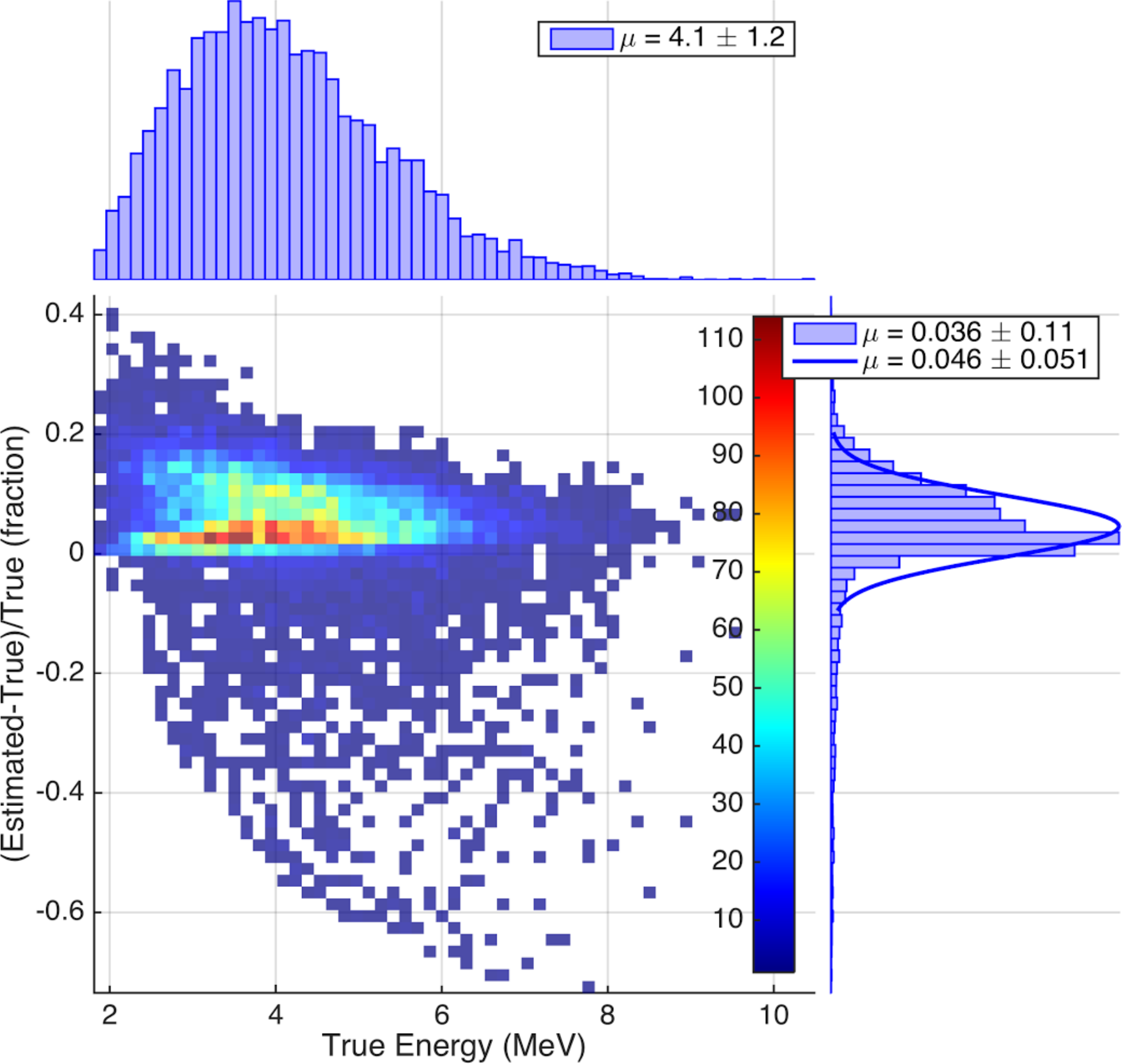}
\caption{Monte Carlo (MC) simulation results showing \nuebar~energy resolution in mTC (y axis) over the reactor antineutrino spectrum (x axis).}
\label{fig:glennh2}
\end{figure}

While the mTC has not yet detected any real world antineutrinos, its performance has been modeled through many GEANT and MATLAB Monte Carlo (MC) simulations. Figure \ref{fig:glennh2} shows the expected \nuebar~energy resolution of the mTC across the 2-10 MeV reactor \nuebar~energy spectrum, which peaks at 3-4 MeV. Our mean energy resolution is about 11\% 1$\sigma$, including outliers in the long tail, or as low as 5\% if outliers are ignored. Most outliers are due to higher energy positrons leaving the detector, resulting in significant under-estimation of their true energy. The prevalence of these occurrences decreases as the wall cuts are expanded.

\begin{figure}[htbp!]
\includegraphics[width=1.0\linewidth]{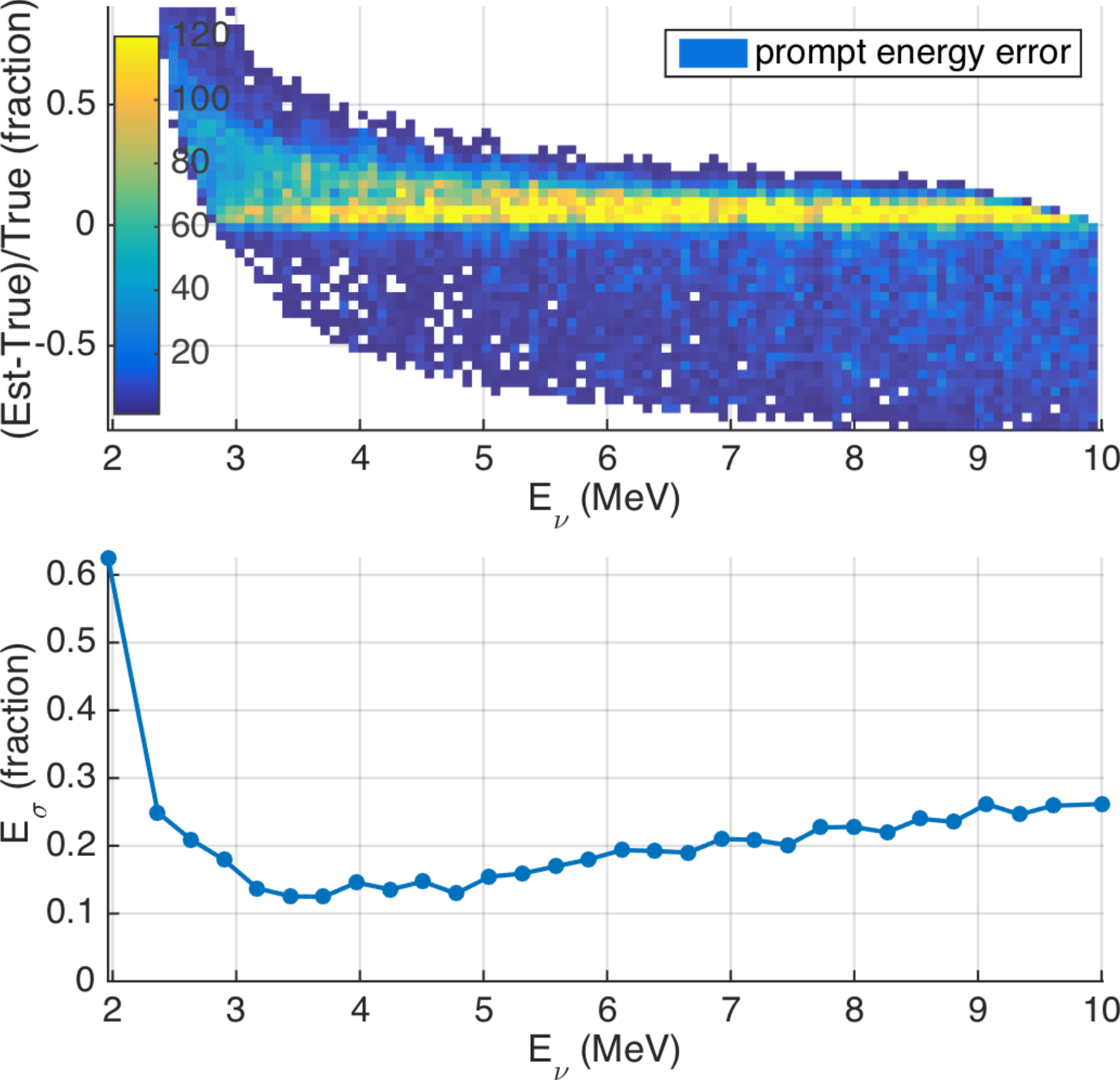}
\caption{Simulated mTC \nuebar~energy resolution vs.  \nuebar~energy.}
\label{fig:glennf3v}
\end{figure}

\begin{figure}[htbp!]
\includegraphics[width=1.0\linewidth]{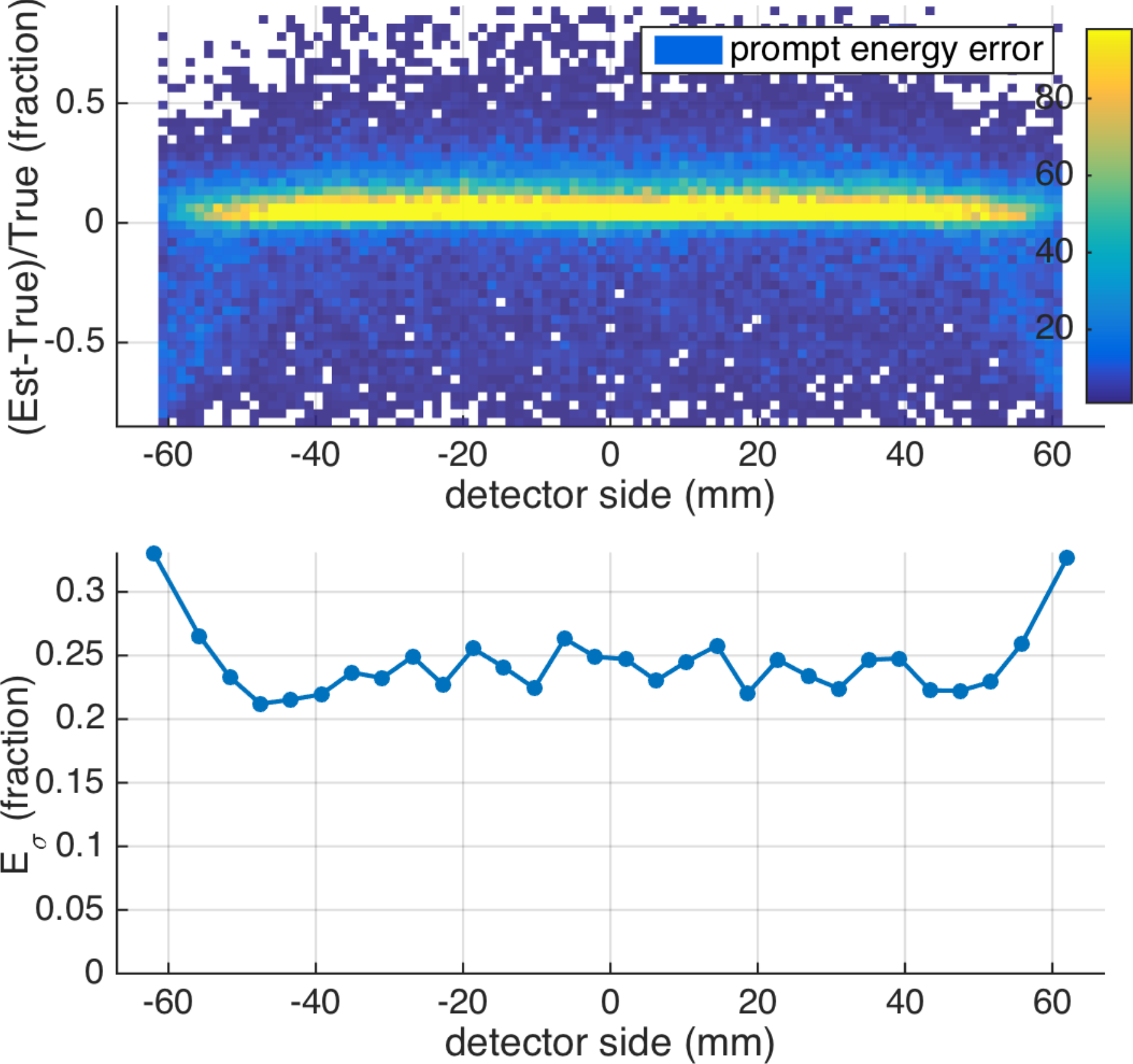}
\caption{Simulated mTC \nuebar~energy resolution vs. vertex location, averaged over \nuebar~energy with a flat input spectrum.  The detector center is at 0 mm, and the detector wall is at 67 mm.}
\label{fig:glennf10v}
\end{figure}

Figure \ref{fig:glennf3v} shows the corresponding energy resolution as a function of \nuebar~energy rather than weighted by the reactor spectrum as in Fig.~\ref{fig:glennh2}. A nice coincidence is seen here: the best energy resolution is enjoyed at the peak of the reactor spectrum, with the resolution suffering at lower energies due to lack of light, and suffering at higher energies due to the high energy positron tracks leaving the detector more frequently.  Figure \ref{fig:glennf10v} shows the same energy resolution values vs. vertex within the detector, indicating that resolution suffers near the detector walls.

\begin{figure}[htbp!]
\includegraphics[width=1.0\linewidth]{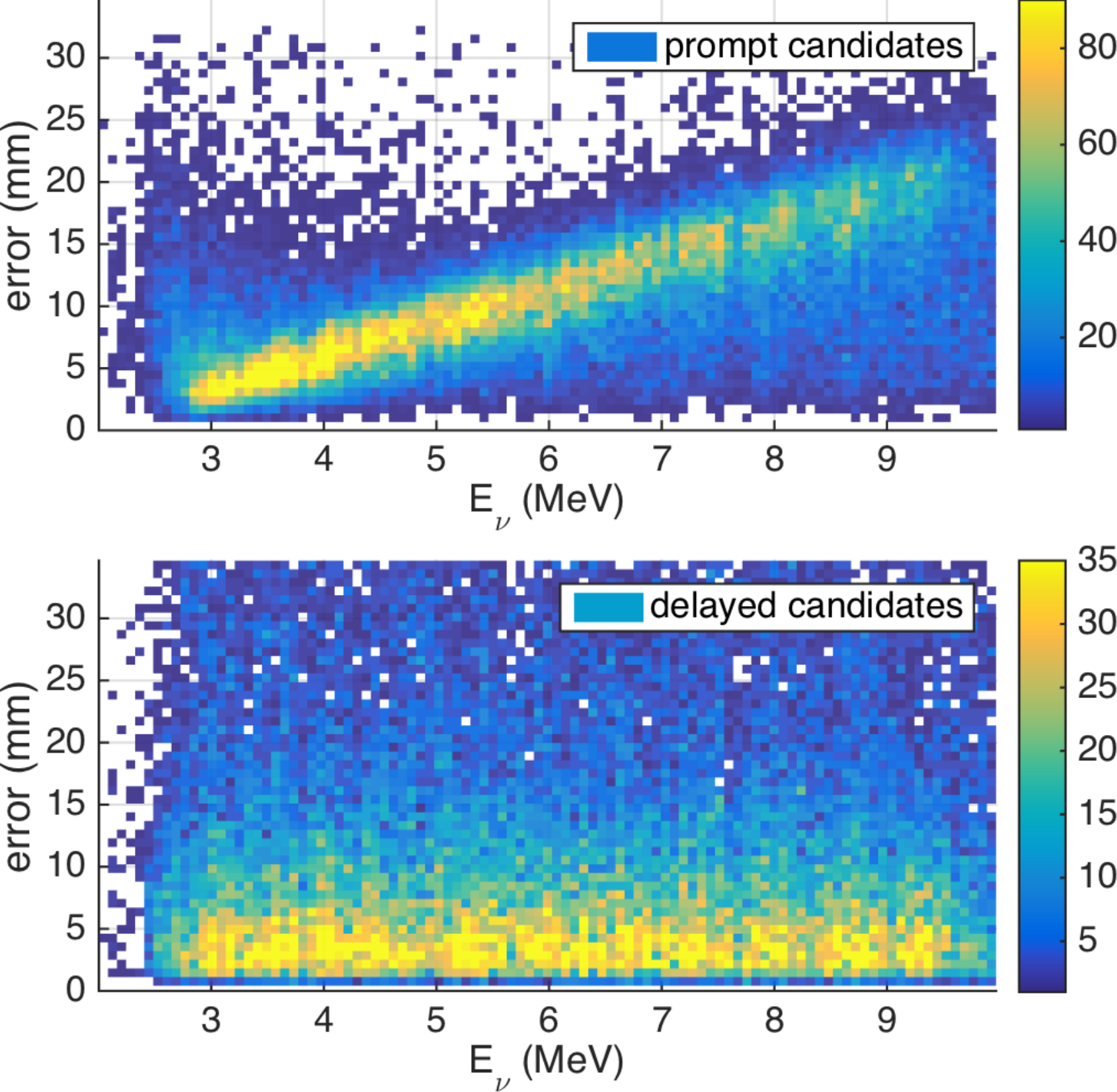}
\caption{Simulated mTC \nuebar~prompt and delayed vertex resolution vs.  \nuebar~energy.}
\label{fig:glennf5v}
\end{figure}

Figure \ref{fig:glennf5v} shows the prompt and delayed vertex resolution as a function of \nuebar~energy. In this context `vertex' means the \nuebar~interaction point for the prompt signal (the start of the $e^+$ track), and the capture location of the neutron for the delayed signal. The prompt vertex fits tend to bias towards the center of the $e^+$ track rather than its start, and both the prompt and delayed vertex location fits are smeared by the spatially dispersed energy depositions of the prompt (2$\times$ 511 keV) and delayed (1$\times$ 470 keV) gammas.

\begin{figure}[htbp!]
\includegraphics[width=1.0\linewidth]{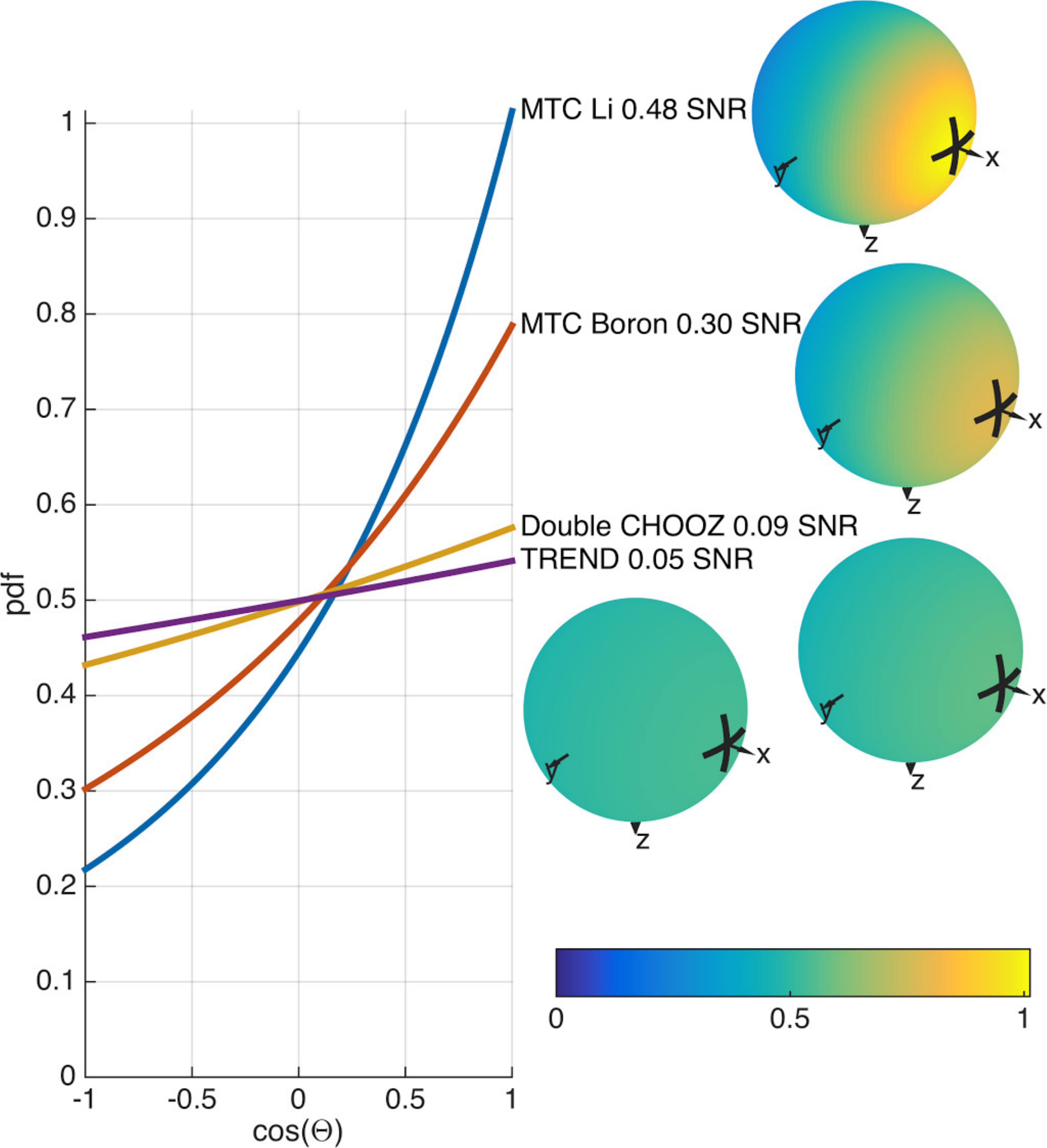}
\caption{Angular \nuebar~resolution comparison between mTC, CHOOZ and TREND. CHOOZ and Double CHOOZ detectors both provide nearly identical angular resolutions as a result of identical near and far detector designs.}
\label{fig:glennf8h}
\end{figure}

Figure \ref{fig:glennf5v} also shows that the prompt vertex resolution suffers at low \nuebar~energies due to lack of light, and at higher \nuebar~energies due to longer $e^+$ tracks (as the center of the track distances itself from its start point). Figure \ref{fig:glennf8h} shows Monte Carlo \nuebar~angle reconstructions in the mTC and puts the mTC angular resolution in context by comparison with the CHOOZ detectors\cite{Abe:2014bwa} and hypothetical 138 kT TREND detector.\cite{Jocher:2013gta}

Figure \ref{fig:glennf8h} shows the angle error distributions of mTC, CHOOZ, and TREND over the -1 to 1 cos$(\theta)$ range, where cos$(\theta)=1$ corresponds to zero error and $cos(\theta)=-1$ corresponds to 180$^\circ$ of angle error. The color maps on the unit sphere represent several thousand reconstructions of Monte Carlo events, and serve as a more intuitive measure of how well the mTC reconstructs direction. 
Though the mTC hypothetically exceeds the \nuebar~angular resolution of the CHOOZ detector, they are both in reality extremely poor at directional determination from a single \nuebar, and require great statistics to reduce the uncertainty on any angle fit. 

The angular resolution metric we employ is the vector Signal to Noise Ratio (vector SNR). The vector SNR is the magnitude of the vector mean divided by the standard deviation in any of the 3 dimensions (which should all share similar uncertainties) for a given population of vectors. In our application these vectors are the reconstruction vectors connecting the delayed signal vertices to the prompt signal vertices. Such a group of vectors should, with some uncertainty, point back towards the \nuebar~source.

We use this metric rather than the more common angle 1$\sigma$ because the uncertainty is so great as to wrap completely around the sphere, rendering simpler 1-dimensional methods meaningless. An alternative metric for directional statistics is the von Mises--Fisher distribution, which provides a `concentration parameter' that increases as the angular distribution decreases.

In the mTC, our mean reconstruction vector (from delayed vertex to prompt vertex) is 10 mm long, and the 1$\sigma$ standard deviation of these vectors is 32 mm, giving us a vector SNR of 10 mm / 32 mm = 0.3. In the CHOOZ \nuebar~detector, the mean reconstruction vector is 17 mm long with a 190 mm $1\sigma$ uncertainty about each axis, producing an SNR of 17 mm / 190 mm = 0.09. The simulated TREND SNR is 0.05. Figure \ref{fig:glennf8h} shows the angular distributions for these 3 detectors plotted over the -1 to 1 cos$(\theta)$ range, as well as wrapped around a unit sphere on a common colormap. Also, a hypothetical mTC-detector is shown with 1.5\%~$\nucl{6}{Li}$-loaded plastic scintillator.

Per Fig.~\ref{fig:glennf8h}, the chances of reconstructing a \nuebar~as originating from the correct hemisphere (i.e. forward or backward) is 62\% in the mTC, 54\% in CHOOZ, and 52\% in TREND. These values are obtained by simply integrating the 0-1 cos$(\theta)$ probabilities in Figure \ref{fig:glennf8h}.

The analytical equation defining the vector SNR distributions shown in Fig.~\ref{fig:glennf8h} is Equation 44 in Jocher {\it et al.}\cite{Jocher:2013gta} This equation defines a proper analytical PDF (one that normalizes to unity) over the -1 to 1 cos$(\theta)$ range, and was derived specifically for the purpose of describing \nuebar~directional resolution. We could not find evidence of its use previously in the field of directional statistics.

\section{Conclusions}
The compact size of the miniTimeCube gives it the potential for many novel measurements. 
Preliminary results indicate the mTC should attain a 1:1 \nuebar~signal-to-noise ratio inside the shielding cave at the NIST reactor site.  The mTC is designed to exploit fast timing for event reconstruction. 
While the fiducial volume may be smaller than desirable in certain aspects, the sizing, coupled with the fast $\sim$ 100~ps electronics and high spatial resolution enable high levels of imaging and reconstruction not attainable by larger detectors. Our modeled antineutrino vertex and angular resolutions (10 mm vector and 32 mm of noise) indicate that the mTC should enjoy $\sim$3 times better directional resolution than existing large \nuebar~detectors.

mTC energy resolution, at 11\% (or 5\% without outliers), is on par with other large antineutrino detectors, despite the challenges faced by uncertain gamma energy deposition from one IBD event to the next. The efficiency of the detector, at 30\%, may be improved significantly by the addition of higher levels of neutron capture doping material,
 which would solve many of the current problems with neutron retention in time and space.

One could scale the mTC concept up to a larger detector --- as is planned with \mbox{NuLat} --- or build a networked array of such small detectors. These could be used to perform in-depth studies ranging from neutrino oscillation with novel detector arrangements at very short baselines or to explore nuclear security applications.
Upgrades to the mTC are ongoing, with continuing calibration, electronics improvements, and reactor tests planned. Ultimately we believe the mTC provides exciting opportunities for fast timing exploitation, and we look forward to publishing future results as they become available.

\begin{acknowledgments}
We gratefully acknowledge the funding for the mTC project provided by \NGA, the U.S. Department of Energy HEP, the National Science Foundation, and the University of Hawaii.

We also acknowledge the support of the NIST, US Department of Commerce, in providing support for facilities used in this work.

We would like to thank NIST personnel for hosting our detector and making the mTC lab at NIST Center for Neutron Research our second home,  
the UH Instrumentation Development Laboratory team for their commitment to improving the electronics, 
NGA for providing guidance, 
Integrity Applications Incorporated,
the UH Physics Department workshop team,
the UH Applied Research Laboratory,
the UH Physics Department IT specialists,
Photonis and WIENER's technical support teams,
Prof. Stephen Dye from Hawaii Pacific University,
UH project manager Andrew Druetzler,
Sharon Messina at the NGA for help with graphic design, 
graduate student Stefanie Smith, undergraduate students Andrew Carpenter and Yenmy Truong who were involved in the project in its early stages, and NIST summer high school student Jason Siegel who worked on the neutron background simulations for the mTC's shielding.

\end{acknowledgments}

\section*{Disclaimer}
Certain trade names and company products are mentioned in the text or identified in illustrations in order to adequately specify the experimental procedure and equipment used. In no case does such identification imply recommendation or endorsement by the National Institute of Standards and Technology, nor does it imply that the products are necessarily the best available for the purpose.

\section{References}\bibliography{manuscript.bib}

\end{document}